\documentclass[english,reprint,tightenlines,eqsecnum,aps,prd,nofootinbib]{revtex4}
\usepackage[T1]{fontenc}
\usepackage[latin9]{inputenc}
\usepackage{amsmath}
\usepackage{amssymb}
\usepackage{stmaryrd}
\usepackage{graphicx}
\usepackage{esint}

\makeatletter
\@ifundefined{textcolor}{}
{%
 \definecolor{BLACK}{gray}{0}
 \definecolor{WHITE}{gray}{1}
 \definecolor{RED}{rgb}{1,0,0}
 \definecolor{GREEN}{rgb}{0,1,0}
 \definecolor{BLUE}{rgb}{0,0,1}
 \definecolor{CYAN}{cmyk}{1,0,0,0}
 \definecolor{MAGENTA}{cmyk}{0,1,0,0}
 \definecolor{YELLOW}{cmyk}{0,0,1,0}
}

\makeatother

\usepackage{babel}
\begin{document}

\title{Quantum effects near the Cauchy horizon of a Reissner-Nordström black
hole}

\author{Orr Sela}

\address{Department of physics, Technion-Israel Institute of Technology, Haifa
32000, Israel}
\begin{abstract}
We consider a massless, minimally-coupled quantum scalar field on
a Reissner-Nordström black hole background, and we study the leading
asymptotic behavior of the expectation value of the stress energy
tensor operator $\langle\hat{T}_{\mu\nu}\rangle_{ren}$ and of $\langle\hat{\Phi}^{2}\rangle_{ren}$
near the inner horizon, in both the Unruh and the Hartle-Hawking quantum
states. We find that the coefficients of the expected leading-order
divergences of these expectation values vanish, indicating that the
modifications of the classical geometry due to quantum vacuum effects
might be weaker than expected. In addition, we calculate the leading-order
divergences of $\langle\hat{T}_{\mu\nu}\rangle_{ren}$ and of $\langle\hat{\Phi}^{2}\rangle_{ren}$
in the Boulware state near the outer (event) horizon, and we obtain
analytical expressions that correspond to previous numerical results.
\end{abstract}
\maketitle

\section{Introduction}

The classical Einstein\textquoteright s field equation of general
relativity admits black hole (BH) solutions with internal structures
that possess exotic features such as naked singularities, Cauchy horizons
and bridges to other universes. However, this is an ideal picture
obtained when considering a highly symmetric, isolated black hole.
Two well known solutions that have this kind of nontrivial internal
structure are the Reissner-Nordström (RN) solution, describing a spherically
symmetric, electrically charged BH, and the Kerr solution, describing
a stationary, rotating and axially-symmetric BH. 

As it turns out, when one adds (classical) external matter and perturbations
to the picture, the internal region of these BH solutions is highly
modified and a null curvature singularity is developed in the ingoing
section of the inner horizon \textendash{} the Cauchy horizon. This
singularity has a very interesting nature \textendash{} it is a weak
singularity, meaning that the metric there is continuous but not differentiable.
As a result, an observer can pass through this singularity and experience
a regular physical effect (such as finite tidal forces). This phenomenon
has been shown to occur for both (four dimensional) spherical charged
black holes \cite{cl sing scbh 1,cl sing scbh 2,cl sing scbh 3} and
rotating black holes \cite{cl sing rbh 1,cl sing rbh 2,cl sing rbh 3,cl sing rbh 4}.
Moreover, this picture is qualitatively the same both for a test-field
analysis and for an analysis that takes into account the back-reaction
of the perturbations. Recently, it was shown that even on the outgoing
section of the inner horizon an interesting singularity is developed
\textendash{} a shock-wave singularity \cite{Marolf =000026 Ori,Eilon =000026 Ori}. 

In addition to these classical effects, a very important source for
stress-energy (and as a result for back-reaction) is the quantum nature
of fields. As Hawking \cite{Hawking 1,Hawking 2} showed, considering
quantum fields on a (classical) BH background might result in a significant
effect on the geometry (at least if taken for long enough time) and
change one's qualitative picture of this spacetime. In particular,
he showed that a BH should evaporate through an emission of radiation,
and therefore the entire structure of this BH spacetime turns out
to be very different than the one proposed by the classical picture. 

In this paper, we aim at investigating the stress-energy resulting
from quantum fields in the interior region of a BH with a nontrivial
internal structure (as discussed above). Specifically, we are interested
in the behavior of the stress-energy tensor near the Cauchy horizon,
where it is expected to diverge \cite{Birrell =000026 Davies paper,Hiscock,Ottewill =000026 Winstanley}.
Analyzing this divergence can provide us an important insight into
the possible modification of the interior geometry caused by the quantum
theory. 

We consider the framework of semiclassical general relativity, in
which the gravitational field is treated classically as a curvature
of spacetime, while all the other fields are taken as quantum fields
residing in this background. Therefore, instead of the classical Einstein\textquoteright s
field equation, we now have the following: 
\begin{equation}
G_{\mu\nu}=8\pi\left\langle \hat{T}_{\mu\nu}\right\rangle _{ren},\label{Semi-class field eq}
\end{equation}
where $G_{\mu\nu}$ is the Einstein tensor of the background geometry
and $\langle\hat{T}_{\mu\nu}\rangle_{ren}$ is the renormalized expectation
value of the stress-energy tensor operator associated with the quantum
fields. In Eq. \eqref{Semi-class field eq} and throughout this paper
we adopt standard geometric units $c=G=1$ and the signature $\left(-+++\right)$. 

In Ref. \cite{Birrell =000026 Davies paper}, it was shown that the
renormalized stress-energy tensor (RSET), calculated in the Hartle-Hawking
state on a two dimensional RN black hole background, diverges at the
inner horizon. Then, it was claimed that also in four dimensions the
RSET is expected to diverge at the inner horizon. However, this divergence
was only claimed on general grounds, without any actual calculation. 

In Refs. \cite{Hiscock,Ottewill =000026 Winstanley}, it was shown
that the RSET, calculated in the Unruh state on a four dimensional
RN black hole background, must diverge on at least one of the two
inner horizons. It was also shown that this result applies to slowly
rotating black holes as well under the assumption that some components
of the RSET are continuous functions of $a$, the Kerr angular momentum
parameter. A stronger assumption, that of analyticity in $a$, yields
this divergence for a general value of $a$. This divergence, in turn,
suggests that the classical picture is strongly modified by quantum
vacuum effects associated with the Hawking evaporation process. However,
as in \cite{Birrell =000026 Davies paper}, the form of this divergence
was not found. 

Our goal in this work is to analyze these expected divergences at
the inner horizons of black holes with a nontrivial internal structure.
For this purpose, we consider a simple model with the above mentioned
features: A massless, minimally coupled quantum scalar field on a
RN black hole background. Being massless and minimally coupled, the
scalar field operator $\hat{\Phi}\left(x\right)$ satisfies the d'alembertian
equation: 
\begin{equation}
\square\hat{\Phi}\left(x\right)=0,\label{d'ale phi}
\end{equation}
where the metric used in the calculation of this d'alembertian is
the RN metric. Then, after finding the asymptotic form (near the inner
horizon) of the scalar field modes involved in the expression for
the RSET, we calculate the expected leading-order divergence of this
RSET near the inner horizon and show that its coefficient vanishes.
This, in turn, suggests that the modification of the geometry might
be weaker than expected. Our logic is based on experience from investigating
two dimensional black hole models, where the asymptotic form of the
RSET near the inner horizon determines the modification of the geometry
there, even when back-reaction is taken into account \cite{Amos 2d}. 

An important point in our calculation of the asymptotic form of the
modes near the inner horizon is that the large-$l$ limit is taken,
where $l$ is the usual number appearing in the angular decomposition
of the modes into spherical harmonics. That is, we assume that the
leading-order behavior of $\langle\hat{T}_{\mu\nu}\rangle_{ren}$
and $\langle\hat{\Phi}^{2}\rangle_{ren}$ near the inner horizon is
determined by the large-$l$ modes. In the external region of the
BH, this assumption turns out to yield the exact asymptotic form of
the RSET near the outer horizon numerically computed in \cite{Anderson,Anderson phi2},
and in the internal region it is consistent with the numerical results
of \cite{Letter} that will be published elsewhere (see below for
more details). For other approximation schemes which have been employed
in different cases than the one considered in this paper, see \cite{Huang}
for an approximation of the RSET of a conformally coupled scalar field
on a RN background, and \cite{Frolov,Anderson 1,Anderson 2,Zannias}
for related discussions. See also \cite{Anderson,Trace anomaly}.

The organization of this paper is as follows. We start in section
II with the preliminaries needed for our analysis. Then, in section
III, and before we turn to the analysis in the interior region of
the BH, we apply our analytical approach to the calculation of the
leading-order divergence of the RSET (and of $\langle\hat{\Phi}^{2}\rangle_{ren}$,
where $\hat{\Phi}$ is the scalar field operator)\emph{ }in Boulware
state upon approaching the outer (event) horizon from outside of the
BH. Later, in section IV, we review some useful results from \cite{Group},
where the two-point function in the interior region was expressed
in terms of a radial function (or alternatively, in terms of some
inner modes) that can be analytically calculated in the asymptotic
region near the inner horizon. Then, in section V, we find this asymptotic
form of the radial function and use it to calculate in section VI
the leading divergence of $\langle\hat{\Phi}^{2}\rangle_{ren}$ and
$\langle\hat{T}_{\mu\nu}\rangle_{ren}$ near the inner horizon. We
finally conclude in section VII.  

\section{Preliminaries}

\subsection{Coordinate systems and quantum states\label{sec:Quantum-states}}

In this paper we consider the Reissner-Nordström spacetime, which
in the standard Schwarzschild coordinates has the following metric:
\[
ds^{2}=-\left(1-\frac{2M}{r}+\frac{Q^{2}}{r^{2}}\right)dt^{2}+\left(1-\frac{2M}{r}+\frac{Q^{2}}{r^{2}}\right)^{-1}dr^{2}+r^{2}\left(d\theta^{2}+\sin^{2}\theta d\varphi^{2}\right).
\]
We define the various choices of coordinates on this space following
\cite{Group}. First, we define the tortoise coordinate, $r_{*}$,
using the standard relation
\[
\frac{dr}{dr_{*}}=1-\frac{2M}{r}+\frac{Q^{2}}{r^{2}}.
\]
We use this relation to define $r_{*}$ both in the interior and the
exterior regions of the BH. More explicitly, we choose the integration
constants such that $r_{*}$ is given by 
\begin{equation}
r_{*}=r+\frac{1}{2\kappa_{+}}\ln\left(\frac{\left|r-r_{+}\right|}{r_{+}-r_{-}}\right)-\frac{1}{2\kappa_{-}}\ln\left(\frac{\left|r-r_{-}\right|}{r_{+}-r_{-}}\right)\label{rs definition}
\end{equation}
in both regions, where $\kappa_{\pm}$ are the surface gravities of
the BH corresponding to the inner and outer horizons, and are defined
as 
\[
\kappa_{\pm}=\frac{r_{+}-r_{-}}{2r_{\pm}^{2}}.
\]
Notice that both $\kappa_{+}$ and $\kappa_{-}$ are chosen to be
positive. Using Eq. \eqref{rs definition}, it is easy to see that
the outer horizon (at $r=r_{+}$) corresponds to $r_{*}\rightarrow-\infty$
(both for $r_{*}$ defined in the exterior region and for that defined
in the interior) and the inner horizon (at $r=r_{-}$) to $r_{*}\rightarrow\infty$. 

\begin{figure}
\begin{centering}
\includegraphics[scale=0.6]{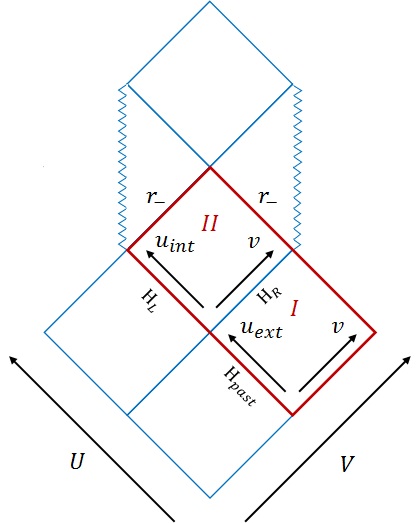}
\par\end{centering}
\caption{Penrose diagram of Reissner-Nordström spacetime. In the exterior region,
region $I$ in the figure, we use the exterior Eddington-Finkelstein
coordinates, while in the interior, region $II$ in the figure, we
use the interior Eddington-Finkelstein coordinates. In addition, the
Kruskal coordinate system is shown and is defined in both regions
$I$ and $II$. The red-framed area denotes the region in the eternal
Reissner-Nordström spacetime which concerns this paper, i.e. regions
$I$ and $II$. \label{fig:Penrose diag RN}}
\end{figure}

Next, we define the Eddington-Finkelstein coordinates in the exterior
region by
\[
u_{\mathrm{ext}}=t-r_{*}\quad,\quad v=t+r_{*},
\]
while in the interior region by 
\begin{equation}
u_{\mathrm{int}}=r_{*}-t\quad,\quad v=r_{*}+t,\label{v =000026 u_int}
\end{equation}
see Fig. \eqref{fig:Penrose diag RN}. The Kruskal coordinates corresponding
to the event horizon, $r=r_{+}$, are defined in terms of the exterior
and interior Eddington-Finkelstein coordinates by 
\[
U\left(u_{\mathrm{ext}}\right)=-\frac{1}{\kappa_{+}}\exp\left(-\kappa_{+}u_{\mathrm{ext}}\right),\,\quad U\left(u_{\mathrm{int}}\right)=\frac{1}{\kappa_{+}}\exp\left(\kappa_{+}u_{\mathrm{int}}\right),\,\quad V\left(v\right)=\frac{1}{\kappa_{+}}\exp\left(\kappa_{+}v\right).
\]
Note that we are interested in regions I and II of Fig. \eqref{fig:Penrose diag RN}
{[}i.e. the region $\left(-\infty<U<\infty\,,\,V>0\right)${]}, in
which the coordinate $v$ is well defined and so we do not need to
introduce different coordinates, $v_{\mathrm{ext}}$ and $v_{\mathrm{int}}$,
for the exterior and interior regions. 

We make the following notations: $H_{\mathrm{past}}$ denotes the
past horizon {[}i.e. the region $\left(U<0\,,\,V=0\right)${]}, PNI
denotes past-null-infinity {[}i.e. $\left(U=-\infty\,,\,V>0\right)${]},
$H_{L}$ is the region $\left(U>0\,,\,V=0\right)$ and $H_{R}$ is
the region $\left(U=0\,,\,V>0\right)$, see Fig. \eqref{fig:Penrose diag RN}.
We call $H_{L}$ and $H_{R}$ the ``left event horizon'' and ``right
event horizon'', respectively. 

Let us now define the three quantum states that we consider in this
paper. Again, we follow the notations of \cite{Group}. We begin with
defining the Boulware state \cite{Boulware} which is defined in the
exterior region of the BH. In order to do so, we decompose our (massless,
minimally coupled) scalar field operator $\hat{\Phi}\left(x\right)$
in the exterior region using two independent sets of modes $f_{\omega lm}^{\Lambda}\left(x\right)$,
known as the Boulware modes, where $\varLambda$ denotes ``in''
and ``up''. The decomposition is given by 
\begin{equation}
\hat{\Phi}\left(x\right)=\int d\omega\sum_{\Lambda,l,m}\left[f_{\omega lm}^{\Lambda}\left(x\right)\hat{a}_{\omega lm}^{\Lambda}+f_{\omega lm}^{\Lambda*}\left(x\right)\hat{a}_{\omega lm}^{\Lambda\dagger}\right]\label{operator decom}
\end{equation}
and the Boulware modes $f_{\omega lm}^{\Lambda}$ are defined as follows.
First, these modes are solutions of the d'alembertian equation satisfied
by the scalar field operator $\hat{\Phi}\left(x\right)$ {[}see Eq.
\eqref{d'ale phi}{]}, i.e. 
\begin{equation}
\square f_{\omega lm}^{\Lambda}\left(x\right)=0.\label{field eq modes}
\end{equation}
Exploiting the spherical symmetry, we can decompose $f_{\omega lm}^{\Lambda}$
as 
\begin{equation}
f_{\omega lm}^{\Lambda}\left(x\right)=\left|\omega\right|^{-1/2}C_{lm}\left(x\right)\tilde{f}_{\omega l}^{\Lambda}\left(x\right)\label{decom f}
\end{equation}
where 
\begin{equation}
C_{lm}\left(x\right)=\left(4\pi\right)^{-1/2}\frac{1}{r}Y_{lm}\left(\theta,\varphi\right),\label{Ct}
\end{equation}
and get a two-dimensional wave equation for $\tilde{f}_{\omega l}^{\Lambda}$:
\begin{equation}
\tilde{f}_{,r_{*}r_{*}}^{\Lambda}-\tilde{f}_{,tt}^{\Lambda}=V_{l}\left(r\right)\tilde{f}^{\Lambda}\label{Wave Eq ft}
\end{equation}
where 
\begin{equation}
V_{l}\left(r\right)=\left(1-\frac{2M}{r}+\frac{Q^{2}}{r^{2}}\right)\left[\frac{l\left(l+1\right)}{r^{2}}+\frac{2M}{r^{3}}-\frac{2Q^{2}}{r^{4}}\right].\label{Potential}
\end{equation}
The Boulware modes are then defined by demanding that $\tilde{f}_{\omega l}^{\Lambda}$
are the solutions of Eq. \eqref{Wave Eq ft} that satisfy the following
initial conditions: 

\begin{equation}
\tilde{f}_{\omega l}^{\mathrm{in}}=\left\{ \begin{array}{c}
0\quad,\quad H_{\mathrm{past}}\\
e^{-i\omega v}\quad,\quad\mathrm{PNI}
\end{array}\right.,\label{i.c in B mode}
\end{equation}
\begin{equation}
\tilde{f}_{\omega l}^{\mathrm{up}}=\left\{ \begin{array}{c}
e^{-i\omega u_{\mathrm{ext}}}\quad,\quad H_{\mathrm{past}}\\
0\quad,\quad\mathrm{PNI}
\end{array}\right..\label{i.c up B mode}
\end{equation}
The Boulware state, $\left|0\right\rangle _{B}$, is then defined
by 
\begin{equation}
\hat{a}_{\omega lm}^{\Lambda}\left|0\right\rangle _{B}=0.\label{State}
\end{equation}
Note that these modes and quantum state are defined in the exterior
region of the BH {[}region I of Fig. \eqref{fig:Penrose diag RN}{]}.
This state corresponds to the familiar concept of an empty state at
spatial infinity, in the sense that the expectation value of the stress-energy
tensor (in asymptotically Lorentzian coordinate system) in this state
goes to zero at large radii \cite{Candelas}. Moreover, this expectation
value, evaluated in a freely falling frame, diverges at the event
horizons. 

An alternative and natural name for the Boulware modes that we will
use in this paper is the ``outer Eddington-Finkelstein modes''.
Analogously to the definition of these modes, we can also define ``inner
Eddington-Finkelstein modes''. We shall use a similar notation for
these modes, $f_{\omega lm}^{\Lambda}$ and $\tilde{f}_{\omega l}^{\Lambda}$
{[}related by Eq. \eqref{decom f}{]}, but with $\Lambda$ denoting
``right'' ($R$) and ``left'' ($L$) instead of ``in'' and ``up''.
These modes are defined in the interior region of the BH {[}region
II of Fig. \eqref{fig:Penrose diag RN}{]} by the following initial
conditions on the left and right event horizons: 

\begin{equation}
\tilde{f}_{\omega l}^{\mathrm{L}}=\left\{ \begin{array}{c}
e^{-i\omega u_{\mathrm{int}}}\quad,\quad H_{L}\\
0\quad,\quad H_{R}
\end{array}\right.,\label{b.c L mode}
\end{equation}
\begin{equation}
\tilde{f}_{\omega l}^{\mathrm{R}}=\left\{ \begin{array}{c}
0\quad,\quad H_{L}\\
e^{-i\omega v}\quad,\quad H_{R}
\end{array}\right..\label{b.c R mode}
\end{equation}
We will use these modes repeatedly later in the paper. Note, however,
that we do not involve these modes in a definition of a quantum state. 

Let us now turn to the definition of the Unruh state \cite{Unruh}.
The field operator is decomposed as in Eq. \eqref{operator decom}
and the modes {[}satisfying Eq. \eqref{field eq modes}{]} as in Eq.
\eqref{decom f}. Everything remains the same except for the initial
conditions for the modes $\tilde{g}_{\omega l}^{\Lambda}$ (where
we use the letter $g$ for the Unruh modes) that now take the form
\[
\tilde{g}_{\omega l}^{\mathrm{up}}=\left\{ \begin{array}{c}
e^{-i\omega U}\quad,\quad H_{\mathrm{past}}\cup H_{L}\\
0\quad,\quad\mathrm{\mathrm{PNI}}
\end{array}\right.
\]
and 
\[
\tilde{g}_{\omega l}^{\mathrm{in}}=\left\{ \begin{array}{c}
0\quad,\quad H_{\mathrm{past}}\cup H_{L}\\
e^{-i\omega v}\quad,\quad\mathrm{\mathrm{PNI}}
\end{array}\right..
\]
Using the decomposition \eqref{operator decom} but with the Unruh
modes defined above, the Unruh state is defined as in Eq. \eqref{State}.
Explicitly, if we decompose the scalar field operator $\hat{\Phi}\left(x\right)$
as 
\begin{equation}
\hat{\Phi}\left(x\right)=\int d\omega\sum_{\Lambda,l,m}\left[g_{\omega lm}^{\Lambda}\left(x\right)\hat{b}_{\omega lm}^{\Lambda}+g_{\omega lm}^{\Lambda*}\left(x\right)\hat{b}_{\omega lm}^{\Lambda\dagger}\right],\label{operator decom Unruh}
\end{equation}
the Unruh state is defined by 
\begin{equation}
\hat{b}_{\omega lm}^{\Lambda}\left|0\right\rangle _{U}=0.\label{Unruh state}
\end{equation}

Notice that the Unruh modes (and quantum state) are defined both in
the interior and the exterior regions of the BH {[}regions I and II
of Fig. \eqref{fig:Penrose diag RN}{]}. 

In Unruh state, the expectation value of the stress-energy tensor
corresponds to the Hawking radiation at infinity and it is regular,
in a freely falling frame, on the future event horizon but not on
the past horizon \cite{Candelas}. 

Finally, the third quantum state that we consider is the Hartle-Hawking
state \cite{Hartle =000026 Hawking}. It is usually defined by an
analytic continuation to the Euclidean sector, but we will mainly
be interested in the mode structure of the state, as we used above
for the Unruh and Boulware states. In fact, we shall be interested
in known mode-sum expressions for various expectation values at the
Hartle-Hawking state and will not need to use the precise form of
the modes themselves, see below. The expectation value of the stress-energy
tensor in Hartle-Hawking state corresponds to a thermal bath of radiation
at infinity and it is regular, in a freely falling frame, on both
of the event horizons \cite{Candelas}. 

This state is denoted by $\left|0\right\rangle _{H}$. 

\subsection{Regularization and Renormalization\label{subsec:Reg=000026Ren}}

In this paper, we consider two kinds of composite operators, $\hat{\Phi}^{2}\left(x\right)$
and $\hat{T}_{\mu\nu}\left(x\right)$, which are quadratic in the
scalar field operator and its derivatives. As is well known, composite
operators formally contain divergences and need to be renormalized
in order to yield a well-defined quantity. In this paper, we follow
the renormalization procedure initiated by Candelas in \cite{Candelas}
and recently further analyzed (and generalized to a much larger extent)
in \cite{Adam phi2 t,Adam phi2 theta,Adam T 1,Adam T 2}. This procedure
is based on the so called \emph{point-splitting} method. We begin
this subsection with briefly reviewing the point-splitting method
and then continue with a description of the pragmatic renormalization
procedure used in this paper. 

\subsubsection{Point-Splitting\label{Point-Splitting}}

When dealing with composite operators which are quadratic in the field
operator and its derivatives, an intuitive way to regularize these
operators is to ``split'' the point $x$ to two distinct points
$x$ and $x'$, and consider the operator which is the product of
the two constituent operators at the two different points. This operator
is obviously well-defined. Next, we can subtract a counterterm that
removes the singularity and take the coincidence limit $x'\rightarrow x$,
thereby obtaining the renormalized operator. This is the so called
point-splitting method. An essential property of the counterterm is
that it is a local geometric quantity that fully captures the singular
piece. In other words, it is independent of the quantum state (and
therefore we can write the renormalization as an operator equation).
Moreover, note that if we consider the vacuum expectation value of
the split operator, we get the standard two-point function. 

In Refs. \cite{Christensen 76,Christensen 78}, Christensen used the
DeWitt-Schwinger expansion of the Feynman Green's function \cite{Schwinger,DeWitt}
(see also \cite{Birrell =000026 Davies}) to obtain the counterterms
needed for the renormalization of $\langle\hat{\Phi}^{2}\rangle$
and $\langle\hat{T}_{\mu\nu}\rangle$ in the point-splitting method.
In the case of $\hat{\Phi}^{2}$, the result is 

\begin{equation}
\left\langle \hat{\Phi}^{2}\left(x\right)\right\rangle _{ren}=\lim_{x'\rightarrow x}\left[\left\langle \hat{\Phi}\left(x\right)\hat{\Phi}\left(x'\right)\right\rangle -G_{DS}\left(x,x'\right)\right],\label{phi2 ren}
\end{equation}
where $G_{DS}\left(x,x'\right)$ is known as the DeWitt-Schwinger
counterterm, which for a scalar field with mass $m$ and coupling
constant $\xi$ takes the form \cite{Christensen 76,Anderson} 
\begin{equation}
G_{DS}\left(x,x'\right)=\frac{1}{8\pi^{2}\sigma}+\frac{m^{2}+\left(\xi-1/6\right)R}{8\pi^{2}}\left[\gamma+\frac{1}{2}\ln\left(\frac{\mu^{2}\left|\sigma\right|}{2}\right)\right]-\frac{m^{2}}{16\pi^{2}}+\frac{1}{96\pi^{2}}R_{\mu\nu}\frac{\sigma^{;\mu}\sigma^{;\nu}}{\sigma}.\label{G DS}
\end{equation}
Here, $\sigma$ is the biscalar of geodetic separation (also known
as Synge's world function) which is equal to one-half of the square
of the geodesic distance between $x$ and $x'$, $\gamma$ is the
Euler constant and $R_{\mu\nu}$ and $R$ are the Ricci tensor and
scalar, respectively. The parameter $\mu$ is not uniquely fixed and
corresponds to the well-known ambiguity in the renormalization procedure. 

For the stress-energy tensor, the procedure is analogous and we have
\begin{equation}
\left\langle \hat{T}_{\mu\nu}\left(x\right)\right\rangle _{ren}=\lim_{x'\rightarrow x}\left[\left\langle \hat{T}_{\mu\nu}\left(x,x'\right)\right\rangle -C_{\mu\nu}^{DS}\left(x,x'\right)\right],\label{T ren}
\end{equation}
where $C_{\mu\nu}^{DS}\left(x,x'\right)$ is the corresponding counterterm,
which again is a local geometric quantity built from $\sigma$ and
the metric. In Christensen's original prescription \cite{Christensen 78},
the expression $\langle\hat{T}_{\mu\nu}\left(x,x'\right)\rangle$,
to which we may call the split stress-energy tensor, involves covariant
derivatives taken at both $x$ and $x'$, along with the bi-vector
of parallel transport which transfers a vector at $x$ to a vector
at $x'$. In this paper, however, we use the alternative form of this
prescription found in \cite{Adam T 2}. In this new form, all the
derivatives in $\langle\hat{T}_{\mu\nu}\left(x,x'\right)\rangle$
are taken at the same point $x$, and the bi-vector of parallel transport
only appears in the new form of the counterterm. This way, the derivation
of the mode-sum expression for $\langle\hat{T}_{\mu\nu}\left(x,x'\right)\rangle$
is much easier and one can naively apply the classical expressions
for the stress-energy tensor for the relevant modes in order to get
this mode-sum. For further discussion and the explicit form of this
new counterterm $C_{\mu\nu}^{DS}\left(x,x'\right)$, see \cite{Adam T 2}. 

\subsubsection{Pragmatic mode-sum renormalization method }

Implementing the point-splitting renormalization is most easily done
using the method introduced in \cite{Candelas,Adam phi2 t,Adam phi2 theta,Adam T 1,Adam T 2}.
In this method, we first use the mode decomposition of the field operator
in order to get a mode-sum expression for the split operator (e.g.
the two-point function in the case of $\langle\hat{\Phi}^{2}\rangle$).
The splitting is done along a direction of a symmetry of the geometry,
i.e. in the direction of a Killing vector. Then, we write the counterterm
as an integral which is of the same kind as one of those used in the
mode-sum expression (this integral is done over a variable which is
conjugate to the symmetry direction coordinate), and subtract the
two. After doing that, we can take the coincidence limit and get the
result (up to some extra regularizations that might be needed). The
surprising thing about this procedure is that one can actually take
the coincidence limit right at the beginning. In other words, one
can write a formal mode-sum expression for the expectation value under
consideration (e.g. $\langle\hat{\Phi}^{2}\rangle$), which is a divergent
quantity, and after doing the same for the counterterm, subtract the
two. The quantity obtained in this way might still be a divergent
one, but after removing these nonphysical divergences one gets the
correct result. 

In this paper we will only be concerned with a simple application
of this renormalization method. Many of the subtleties that arise
in other cases will not occur in our case. The results are derived
using the \emph{$t$-splittin}g variant of the method (see \cite{Adam phi2 t,Adam T 2}),
but they were also confirmed using the \emph{$\theta$-splittin}g
variant (see \cite{Adam phi2 theta}). 

\subsection{The trace of the stress-energy tensor }

When considering a theory with a conformal symmetry, such as a massless,
conformally coupled scalar field, the trace of the stress-energy tensor
operator is a local, geometric quantity (independent of the quantum
state) which is given by \cite{Birrell =000026 Davies}
\begin{equation}
T_{\mathrm{anomaly}}\equiv\frac{1}{2880\pi^{2}}\left(R_{\alpha\beta\gamma\delta}R^{\alpha\beta\gamma\delta}-R_{\alpha\beta}R^{\alpha\beta}+\frac{5}{2}R^{2}+6\boxempty R\right).\label{T_anomaly}
\end{equation}
This result can be generalized to a nonconformal theory as was shown
in \cite{Trace anomaly} and rewritten in a more convenient form for
our analysis in \cite{Group}. In this paper, we consider a massless,
minimally coupled scalar field, for which we have the following result:
\begin{equation}
\left\langle \hat{T}_{\mu}^{\mu}\right\rangle {}_{ren}+\frac{1}{2}\square\left\langle \hat{\Phi}^{2}\right\rangle _{ren}=T_{\mathrm{anomaly}},\label{Trace anomaly}
\end{equation}
relating the expectation values of the two quantities we calculate. 

\section{Expectation value of the stress-energy tensor near the event horizon\label{Exterior}}

Before turning to the calculation of the stress-energy tensor near
the inner horizon, we begin by applying our analytical approach to
the calculation of the leading-order divergence of $\langle\hat{\Phi}^{2}\rangle_{ren}$
and $\langle\hat{T}_{\mu\nu}\rangle_{ren}$\emph{ }in Boulware state
upon approaching the outer (event) horizon from outside of the BH.
We consider a massless, minimally coupled scalar field as our quantum
field. In general, the expectation value of the stress-energy tensor,
evaluated in a freely falling frame, is known to be regular on both
the past and future event (outer) horizons in the case of the Hartle-Hawking
state, it is regular on the future horizon but not on the past horizon
in the case of the Unruh state, and it diverges on both of the horizons
in the case of the Boulware state {[}see the discussion in subSec.
\eqref{sec:Quantum-states} and Ref. \cite{Candelas} for the case
of Schwarzschild spacetime{]}. 

In the present section, in contrast to the analysis in the interior
of the BH (discussed in the rest of the paper), we will not renormalize
the expectation values using the explicit form of the corresponding
counterterms. Instead, we follow a technique presented in \cite{Candelas,Christensen =000026 Fulling}\footnote{Note, however, that in these references the scalar field is conformally
coupled instead of minimally coupled.} and use the fact that the counterterms are geometric quantities independent
of the quantum state, and the fact that the expectation values diverge
at the event horizon most rapidly in Boulware state, in order to obtain
the leading behavior of the renormalized expectation values in this
state using the following subtractions: 
\begin{equation}
\left\langle \hat{\Phi}^{2}\left(x\right)\right\rangle _{B,ren}\cong\left\langle \hat{\Phi}^{2}\left(x\right)\right\rangle _{B,ren}-\left\langle \hat{\Phi}^{2}\left(x\right)\right\rangle _{U,ren}=\left\langle \hat{\Phi}^{2}\left(x\right)\right\rangle _{B}-\left\langle \hat{\Phi}^{2}\left(x\right)\right\rangle _{U},\,\,\,r\rightarrow r_{+}\label{Subtraction phi2}
\end{equation}
and similarly 
\begin{equation}
\left\langle \hat{T}_{\mu\nu}\left(x\right)\right\rangle _{B,ren}\cong\left\langle \hat{T}_{\mu\nu}\left(x\right)\right\rangle _{B,ren}-\left\langle \hat{T}_{\mu\nu}\left(x\right)\right\rangle _{U,ren}=\left\langle \hat{T}_{\mu\nu}\left(x\right)\right\rangle _{B}-\left\langle \hat{T}_{\mu\nu}\left(x\right)\right\rangle _{U},\,\,\,r\rightarrow r_{+},\label{Subtraction T}
\end{equation}
where the subscripts $B$ and $U$ denote the Boulware and Unruh states,
respectively. 

In Refs. \cite{Christensen =000026 Fulling,Candelas}, formal expressions
for $\langle\hat{\Phi}^{2}\rangle$ and $\langle\hat{T}_{\mu\nu}\rangle$
(in the three quantum states discussed here) as mode-sums involving
the Boulware modes {[}see subSec. \eqref{sec:Quantum-states}{]} were
found. Let us quote them here for future reference\footnote{Notice that Refs. \cite{Christensen =000026 Fulling,Candelas} considered
Schwarzschild spacetime and a conformally coupled scalar field, while
we are interested in Reissner-Nordström spacetime and a minimally
coupled scalar field. However, there is no serious qualitative difference
between the two for this analysis (carried out in the exterior region
of the BH), and the expressions map into each other under $\kappa\leftrightarrow\kappa_{+}$,
where $\kappa$ is the surface gravity of the BH, and under a suitable
change in the functional form of the integrand in the mode-sum expression
for the stress-energy tensor. }. If we denote the Boulware modes by $f_{\omega lm}^{\mathrm{up}}$
and $f_{\omega lm}^{\mathrm{in}}$, we have for $\langle\hat{\Phi}^{2}\rangle$
the following expressions: 
\begin{equation}
\left\langle \hat{\Phi}^{2}\left(x\right)\right\rangle _{B}=\intop_{0}^{\infty}d\omega\sum_{l,m}\left[\left|f_{\omega lm}^{\mathrm{up}}\left(x\right)\right|^{2}+\left|f_{\omega lm}^{\mathrm{in}}\left(x\right)\right|^{2}\right],\label{phi2 mode sum B}
\end{equation}
\begin{equation}
\left\langle \hat{\Phi}^{2}\left(x\right)\right\rangle _{U}=\intop_{0}^{\infty}d\omega\sum_{l,m}\left[\coth\left(\frac{\pi\omega}{\kappa_{+}}\right)\left|f_{\omega lm}^{\mathrm{up}}\left(x\right)\right|^{2}+\left|f_{\omega lm}^{\mathrm{in}}\left(x\right)\right|^{2}\right],\label{phi2 mode sum U}
\end{equation}
\begin{equation}
\left\langle \hat{\Phi}^{2}\left(x\right)\right\rangle _{H}=\intop_{0}^{\infty}d\omega\sum_{l,m}\coth\left(\frac{\pi\omega}{\kappa_{+}}\right)\left[\left|f_{\omega lm}^{\mathrm{up}}\left(x\right)\right|^{2}+\left|f_{\omega lm}^{\mathrm{in}}\left(x\right)\right|^{2}\right],\label{phi2 mode sum H}
\end{equation}
where again the subscripts $B$, $U$ and $H$ denote the Boulware,
Unruh and Hartle-Hawking states, respectively. For the stress-energy
tensor $\langle\hat{T}_{\mu\nu}\rangle$, we similarly have 
\begin{equation}
\left\langle \hat{T}_{\mu\nu}\left(x\right)\right\rangle _{B}=\intop_{0}^{\infty}d\omega\sum_{l,m}\left\{ T_{\mu\nu}\left[f_{\omega lm}^{\mathrm{up}}\left(x\right),f_{\omega lm}^{\mathrm{up*}}\left(x\right)\right]+T_{\mu\nu}\left[f_{\omega lm}^{\mathrm{in}}\left(x\right),f_{\omega lm}^{\mathrm{in}*}\left(x\right)\right]\right\} ,\label{T mode sum B}
\end{equation}
\begin{equation}
\left\langle \hat{T}_{\mu\nu}\left(x\right)\right\rangle _{U}=\intop_{0}^{\infty}d\omega\sum_{l,m}\left\{ \coth\left(\frac{\pi\omega}{\kappa_{+}}\right)T_{\mu\nu}\left[f_{\omega lm}^{\mathrm{up}}\left(x\right),f_{\omega lm}^{\mathrm{up*}}\left(x\right)\right]+T_{\mu\nu}\left[f_{\omega lm}^{\mathrm{in}}\left(x\right),f_{\omega lm}^{\mathrm{in}*}\left(x\right)\right]\right\} ,\label{T mode sum U}
\end{equation}
\begin{equation}
\left\langle \hat{T}_{\mu\nu}\left(x\right)\right\rangle _{H}=\intop_{0}^{\infty}d\omega\sum_{l,m}\coth\left(\frac{\pi\omega}{\kappa_{+}}\right)\left\{ T_{\mu\nu}\left[f_{\omega lm}^{\mathrm{up}}\left(x\right),f_{\omega lm}^{\mathrm{up*}}\left(x\right)\right]+T_{\mu\nu}\left[f_{\omega lm}^{\mathrm{in}}\left(x\right),f_{\omega lm}^{\mathrm{in}*}\left(x\right)\right]\right\} ,\label{T mode sum H}
\end{equation}
where 
\[
T_{\mu\nu}\left[f,f^{*}\right]=\frac{1}{2}\left(f_{,\mu}f_{,\nu}^{*}+f_{,\nu}f_{,\mu}^{*}-g_{\mu\nu}g^{\alpha\beta}f_{,\alpha}f_{,\beta}^{*}\right).
\]

Substituting the above expressions for $\langle\hat{\Phi}^{2}\rangle$
and $\langle\hat{T}_{\mu\nu}\rangle$ in the different states into
Eqs. \eqref{Subtraction phi2} and \eqref{Subtraction T}, we get
the following mode-sums for the leading asymptotic behavior of $\langle\hat{\Phi}^{2}\rangle_{B,ren}$
and $\langle\hat{T}_{\mu\nu}\rangle_{B,ren}$ on approaching the outer
(event) horizon from outside of the BH: 
\begin{equation}
\left\langle \hat{\Phi}^{2}\left(x\right)\right\rangle _{B,ren}\cong-2\intop_{0}^{\infty}\frac{d\omega}{e^{2\pi\omega/\kappa_{+}}-1}\sum_{l,m}\left|f_{\omega lm}^{\mathrm{up}}\left(x\right)\right|^{2},\,\,\,r\rightarrow r_{+}\label{phi2 mode-sum rp}
\end{equation}
and 
\begin{equation}
\left\langle \hat{T}_{\mu\nu}\left(x\right)\right\rangle _{B,ren}\cong-2\intop_{0}^{\infty}\frac{d\omega}{e^{2\pi\omega/\kappa_{+}}-1}\sum_{l,m}T_{\mu\nu}\left[f_{\omega lm}^{\mathrm{up}}\left(x\right),f_{\omega lm}^{\mathrm{up*}}\left(x\right)\right],\,\,\,r\rightarrow r_{+}.\label{T mode-sum rp}
\end{equation}

Next, using an asymptotic calculation of the ``up'' Boulware modes
$f_{\omega lm}^{\mathrm{up}}$ near the event horizon, we will find
the explicit asymptotic behavior of $\langle\hat{\Phi}^{2}\rangle_{B,ren}$
and $\langle\hat{T}_{\mu\nu}\rangle_{B,ren}$ by substitution of these
modes into Eqs. \eqref{phi2 mode-sum rp} and \eqref{T mode-sum rp}.
In order to do it, we begin by exploiting the time-translational symmetry
and write $\tilde{f}_{\omega l}^{\Lambda}$ {[}recall that $\tilde{f}_{\omega l}^{\Lambda}$
is related to $f_{\omega lm}^{\Lambda}$ via Eq. \eqref{decom f}{]}
as a product of a simple time-dependent part (common to the ``up''
and ``in'' modes) and a radial function as follows: 
\begin{equation}
\tilde{f}_{\omega l}^{\Lambda}\left(r,t\right)=e^{-i\omega t}\varPsi_{\omega l}^{\varLambda}\left(r\right).\label{decom ft}
\end{equation}
By substituting into Eq. \eqref{Wave Eq ft}, we get the following
equation for the radial functions: 
\begin{equation}
\varPsi_{\omega l}^{\varLambda}{}_{,r_{*}r_{*}}+\left(\omega^{2}-V_{l}\left(r\right)\right)\varPsi_{\omega l}^{\varLambda}=0.\label{Radial eq}
\end{equation}
The asymptotic forms of these functions are easily determined from
those of $\tilde{f}_{\omega l}^{\Lambda}$ {[}given in Eqs. \eqref{i.c in B mode}
and \eqref{i.c up B mode}{]} and are given by 
\[
\varPsi_{\omega l}^{\mathrm{in}}\left(r\right)\cong\left\{ \begin{array}{c}
\tau_{\omega l}^{\mathrm{in}}e^{-i\omega r_{*}}\quad,\quad r_{*}\rightarrow-\infty\\
e^{-i\omega r_{*}}+\rho_{\omega l}^{\mathrm{in}}e^{i\omega r_{*}}\quad,\quad r_{*}\rightarrow\infty
\end{array}\right.
\]
and 

\[
\varPsi_{\omega l}^{\mathrm{up}}\left(r\right)\cong\left\{ \begin{array}{c}
e^{i\omega r_{*}}+\rho_{\omega l}^{\mathrm{up}}e^{-i\omega r_{*}}\quad,\quad r_{*}\rightarrow-\infty\\
\tau_{\omega l}^{\mathrm{up}}e^{i\omega r_{*}}\quad,\quad r_{*}\rightarrow\infty
\end{array}\right.,
\]
where $\rho_{\omega l}^{\Lambda}$ and $\tau_{\omega l}^{\varLambda}$
are the reflection and transmission coefficients (corresponding to
the mode $\Lambda$), respectively. 

Now, since we want to find the asymptotic form of $f_{\omega lm}^{\mathrm{up}}$
near the event horizon, we concentrate on finding the asymptotic form
of $\varPsi_{\omega l}^{\mathrm{up}}\left(r\right)$ near $r_{*}\rightarrow-\infty$.
For that, we expand the potential \eqref{Potential} near $r=r_{+}$
(or $r_{*}\rightarrow-\infty$) and get the asymptotic form 
\begin{equation}
V_{l}\left(r\right)\cong C_{+}\left(r-r_{+}\right)\cong C_{+}\left(r_{+}-r_{-}\right)\exp\left(-2\kappa_{+}r_{+}\right)\exp\left(2\kappa_{+}r_{*}\right),\,\,\,r_{*}\rightarrow-\infty,\label{V exp rp}
\end{equation}
where
\[
C_{+}=\frac{r_{+}-r_{-}}{r_{+}^{4}}\left(l^{2}+l+1-\frac{r_{-}}{r_{+}}\right).
\]
 Substituting this potential into Eq. \eqref{Radial eq}, we find
the following asymptotic form of the radial equation near $r=r_{+}$:
\begin{equation}
R_{\omega l}{}_{,r_{*}r_{*}}+\left[\omega^{2}-C_{+}\left(r_{+}-r_{-}\right)\exp\left(-2\kappa_{+}r_{+}\right)\exp\left(2\kappa_{+}r_{*}\right)\right]R_{\omega l}=0,\label{Radial eq asym}
\end{equation}
where $R_{\omega l}\left(r\right)$ denotes the asymptotic form of
$\varPsi_{\omega l}^{\mathrm{up}}\left(r\right)$ near $r=r_{+}$.
Notice that the boundary condition (or asymptotic form) of $R_{\omega l}\left(r\right)$
at $r_{*}\rightarrow\infty$ is different from that of $\varPsi_{\omega l}^{\mathrm{up}}\left(r\right)$,
because we are considering an exponential potential in Eq. \eqref{Radial eq asym}.
We are looking for a solution $R_{\omega l}\left(r\right)$ that decays
exponentially at $r_{*}\rightarrow\infty$ and takes the form $e^{i\omega r_{*}}+\rho_{\omega l}^{\mathrm{up}}e^{-i\omega r_{*}}$
at $r_{*}\rightarrow-\infty$. The solution that satisfies these conditions
is 
\[
R_{\omega l}\left(r_{*}\right)=\left(\frac{C_{+}r_{+}^{4}}{r_{+}-r_{-}}\right)^{-i\omega/\left(2\kappa_{+}\right)}\frac{2\exp\left(i\omega r_{+}\right)}{\varGamma\left(-i\omega/\kappa_{+}\right)}K_{i\omega/\kappa_{+}}\left(2r_{+}^{2}\sqrt{\frac{C_{+}}{r_{+}-r_{-}}}e^{\kappa_{+}\left(r_{*}-r_{+}\right)}\right),
\]
where $K_{\alpha}\left(z\right)$ is a modified Bessel function of
the second kind. We can easily write this expression in terms of $r$
by noticing that at $r_{*}\rightarrow-\infty$ we have (recall that
we are only interested in $R_{\omega l}$ in this asymptotic region)
\[
e^{\kappa_{+}\left(r_{*}-r_{+}\right)}\cong\left(\frac{r-r_{+}}{r_{+}-r_{-}}\right)^{1/2},\,\,\,r\rightarrow r_{+}.
\]

In the rest of the calculation, we make the assumption that the leading
divergence we are calculating results from large-$l$ values. In addition
to the match between the analytical expressions thus derived and numerical
results (see below), one can motivate this assumption using the following
heuristic argument. First, we notice that the mode-sums \eqref{phi2 mode-sum rp}
and \eqref{T mode-sum rp} contain the factor $\left(e^{2\pi\omega/\kappa_{+}}-1\right)^{-1}$
which decays exponentially at large $\omega$. As a result, the dominant
contribution to the integral over $\omega$ comes from small-$\omega$
values (up to the scale $\kappa_{+}/2\pi\omega$). Next, since at
large-$l$ values (and fixed $r$) the redial functions $\varPsi_{\omega l}^{\varLambda}$
decay exponentially with $l$ according to the WKB approximation to
the solution of Eq. \eqref{Radial eq} (recall that $\omega$ is small),
the dominant contribution to the sum over $l$ comes from $l$-values
that satisfy {[}see the discussion about the WKB method in subSec.
\eqref{Region II}{]} 
\[
l\leq\frac{r^{2}}{M\sqrt{\left(r-r_{+}\right)\left(r-r_{-}\right)}}\equiv l_{\mathrm{Max}}.
\]
At each $r>r_{+}$, $l_{\mathrm{Max}}$ is finite and so is the mode-sum;
however, at the event horizon ($r=r_{+}$) $l_{\mathrm{Max}}$ diverges
and we get an infinite contribution to the mode-sum. This suggests
that the leading (in $r-r_{+}$) divergence we wish to calculate is
a large-$l$ effect and we assume that it is obtained from the leading
large-$l$ behavior. Therefore, we take this limit of $R_{\omega l}$.
It is more convenient to work with the variable 
\begin{equation}
\tilde{l}=l+\frac{1}{2}\label{lt}
\end{equation}
instead of $l$, and so we write $R_{\omega l}$ in terms of $\tilde{l}$
and take the asymptotic large-$\tilde{l}$ form of it. Using 
\begin{equation}
r_{+}^{2}\sqrt{\frac{C_{+}}{r_{+}-r_{-}}}=\tilde{l}+\mathcal{O}\left(\tilde{l}^{-1}\right)\label{Cp large lt}
\end{equation}
we get (writing the expression in terms of $r$) 
\[
R_{\omega l}\left(r\right)\cong\tilde{l}^{-i\omega/\kappa_{+}}\frac{2\exp\left(i\omega r_{+}\right)}{\varGamma\left(-i\omega/\kappa_{+}\right)}K_{i\omega/\kappa_{+}}\left[2\tilde{l}\left(\frac{r-r_{+}}{r_{+}-r_{-}}\right)^{1/2}\right].
\]
Recalling that as $r_{*}\rightarrow-\infty$ this radial function
should take the form $e^{i\omega r_{*}}+\rho_{\omega l}^{\mathrm{up}}e^{-i\omega r_{*}}$,
we get the following expression for $\rho_{\omega l}^{\mathrm{up}}$
(to leading order in $\tilde{l}$): 
\begin{equation}
\rho_{\omega l}^{\mathrm{up}}\cong\tilde{l}^{-2i\omega/\kappa_{+}}e^{2i\omega r_{+}}\frac{\varGamma\left(i\omega/\kappa_{+}\right)}{\varGamma\left(-i\omega/\kappa_{+}\right)},\label{Rho}
\end{equation}
where $\varGamma\left(x\right)$ is the Euler gamma function. 

Now that we found the asymptotic form of $\varPsi_{\omega l}^{\mathrm{up}}\left(r\right)$
near $r_{+}$ (recall that $\varPsi_{\omega l}^{\mathrm{up}}\left(r\right)\cong R_{\omega l}\left(r\right)$),
we have the asymptotic form of $f_{\omega lm}^{\mathrm{up}}$ using
Eqs. \eqref{decom ft} and \eqref{decom f}. Substituting it to Eqs.
\eqref{phi2 mode-sum rp} and \eqref{T mode-sum rp} yields the leading
asymptotic behavior of $\langle\hat{\Phi}^{2}\rangle_{B,ren}$ and
$\langle\hat{T}_{\mu\nu}\rangle_{B,ren}$ on approaching the outer
(event) horizon of the BH\footnote{In these calculations we took advantage of the fact that the leading
contribution comes from large-$l$ values and switched the summation
over $l$ with an integral.}: 
\[
\left\langle \hat{\Phi}^{2}\left(x\right)\right\rangle _{B,ren}\cong-\frac{\kappa_{+}}{96\pi^{2}\left(r-r_{+}\right)},\,\,\,r\rightarrow r_{+}
\]
and
\begin{equation}
\left\langle \hat{T}_{\nu}^{\mu}\left(x\right)\right\rangle _{B,ren}\cong\frac{\kappa_{+}^{2}}{5760\pi^{2}\left(r-r_{+}\right)^{2}}\left(\begin{array}{cccc}
33 & 0 & 0 & 0\\
0 & -11 & 0 & 0\\
0 & 0 & 19 & 0\\
0 & 0 & 0 & 19
\end{array}\right),\,\,\,r\rightarrow r_{+}.\label{T B ren outside}
\end{equation}
Note that the stress-energy tensor is presented in a mixed form and
in Schwarzschild coordinates. A straightforward calculation reveals
that this expression for the stress-energy tensor satisfies the energy-momentum
conservation equation at leading order. Moreover, these two expressions
for $\langle\hat{\Phi}^{2}\rangle_{B,ren}$ and $\langle\hat{T}_{\mu\nu}\rangle_{B,ren}$
satisfy Eq. \eqref{Trace anomaly} at leading order and correspond
to previous numerical results obtained in \cite{Anderson,Anderson phi2}. 

We obtained that the expectation values of $\hat{\Phi}^{2}$ and $\hat{T}_{\mu\nu}$
in Boulware state diverge at the event horizon. We shall proceed to
analyze the expectation values of these operators in Unruh and Hartle-Hawking
states in the interior region of the BH. 

\section{The two-point function in the interior region of the black hole}

In the previous section, we looked at the exterior region of the BH
and used mode-sum expressions for $\langle\hat{\Phi}^{2}\rangle$
and $\langle\hat{T}_{\mu\nu}\rangle$ {[}see Eqs. \eqref{phi2 mode sum B}-\eqref{T mode sum H}{]}
in terms of the Boulware modes (or outer Eddington-Finkelstein modes)
in order to find their asymptotic behavior on approaching the event
(outer) horizon. The season for using the Boulware modes was the simple
equation satisfied by the radial functions of these modes. Finding
the asymptotic form of these radial functions enabled us to get the
desired asymptotic behavior of $\langle\hat{\Phi}^{2}\rangle$ and
$\langle\hat{T}_{\mu\nu}\rangle$. 

In order to investigate the leading-order behavior of $\langle\hat{\Phi}^{2}\rangle$
and $\langle\hat{T}_{\mu\nu}\rangle$ near the inner horizon, we will
use the same technique. That is, we will use mode-sum expressions
in terms of the inner (instead of outer) Eddington-Finkelstein modes,
and after finding the asymptotic form of their radial functions, we
will get the asymptotic behavior of $\langle\hat{\Phi}^{2}\rangle$
and $\langle\hat{T}_{\mu\nu}\rangle$. It is important to note, however,
that even though the general idea of the calculation is the same as
that of the one carried out in the exterior region, many things are
different. For example, the potential that appears in the radial equation
is no longer positive, a fact that influences the analysis of the
radial functions inside the BH. Moreover, we will not use subtractions
as in Eqs. \eqref{Subtraction phi2} and \eqref{Subtraction T}. Instead,
we will use explicit renormalization using counterterms. 

In this section, we review the form of the mode-sum expression for
the two-point function in the interior region found in Ref. \cite{Group}.
The mode-sum is in terms of the inner Eddington-Finkelstein modes,
defined in Eqs. \eqref{b.c L mode}, \eqref{b.c R mode} and \eqref{decom f}.
After taking the coincidence limit, one gets $\langle\hat{\Phi}^{2}\rangle$.
$\langle\hat{T}_{\mu\nu}\rangle$ can also be obtained after the application
of a certain differential operator\footnote{Note that if one only takes the coincidence limit, the result is a
formal mode-sum expression for $<\hat{\Phi}^{2}>$ and $<\hat{T}_{\mu\nu}>$.
If, on the other hand, one subtracts the corresponding counterterms
in the process, the result is the renormalized forms. }. The focus in \cite{Group} is on the symmetrized form of the two-point
function, which is also known as the Hadamard elementary function
and is defined by 
\[
G^{\left(1\right)}\left(x,x'\right)=\left\langle \left\{ \hat{\Phi}\left(x\right),\hat{\Phi}\left(x'\right)\right\} \right\rangle .
\]
There, it was found that in the Unruh state we have 
\[
G_{U}^{\left(1\right)}\left(x,x'\right)=\sum_{l,m}\intop_{0}^{\infty}d\omega\left[\coth\left(\frac{\pi\omega}{\kappa_{+}}\right)\left\{ f_{\omega lm}^{\mathrm{L}}\left(x\right),f_{\omega lm}^{\mathrm{L}*}\left(x'\right)\right\} +\left(\coth\left(\frac{\pi\omega}{\kappa_{+}}\right)\left|\rho_{\omega l}^{\mathrm{up}}\right|^{2}+\left|\tau_{\omega l}^{\mathrm{up}}\right|^{2}\right)\left\{ f_{\omega lm}^{\mathrm{R}}\left(x\right),f_{\omega lm}^{\mathrm{R}*}\left(x'\right)\right\} \right.
\]
\begin{equation}
\left.+2\mathrm{csch}\left(\frac{\pi\omega}{\kappa_{+}}\right)\mathrm{Re}\left(\rho_{\omega l}^{\mathrm{up}}\left\{ f_{\omega lm}^{\mathrm{R}}\left(x\right),f_{-\omega lm}^{\mathrm{L}*}\left(x'\right)\right\} \right)\right],\label{G Unruh}
\end{equation}
where $\rho_{\omega l}^{\mathrm{up}}$ and $\tau_{\omega l}^{\mathrm{up}}$
are the reflection and transmission coefficients that correspond to
the ``up'' Boulware mode (in the exterior region), respectively.
We define curly brackets acting on functions as a symmetrization with
respect to the arguments of these functions, i.e. 
\[
\left\{ A\left(x\right),B\left(x'\right)\right\} =A\left(x\right)B\left(x'\right)+A\left(x'\right)B\left(x\right).
\]

In the Hartle-Hawking state, the Hadamard function takes the form
\[
G_{H}^{\left(1\right)}\left(x,x'\right)=\sum_{l,m}\intop_{0}^{\infty}d\omega\left[\coth\left(\frac{\pi\omega}{\kappa_{+}}\right)\left(\left\{ f_{\omega lm}^{\mathrm{L}}\left(x\right),f_{\omega lm}^{\mathrm{L}*}\left(x'\right)\right\} +\left\{ f_{\omega lm}^{\mathrm{R}}\left(x\right),f_{\omega lm}^{\mathrm{R}*}\left(x'\right)\right\} \right)\right.
\]
\begin{equation}
\left.+2\mathrm{csch}\left(\frac{\pi\omega}{\kappa_{+}}\right)\mathrm{Re}\left(\rho_{\omega l}^{\mathrm{up}}\left\{ f_{\omega lm}^{\mathrm{R}}\left(x\right),f_{-\omega lm}^{\mathrm{L}*}\left(x'\right)\right\} \right)\right].\label{G HH}
\end{equation}

As mentioned above, from these Hadamard functions one can obtain $\langle\hat{\Phi}^{2}\rangle$
and $\langle\hat{T}_{\mu\nu}\rangle$ in the respective states. 

Next, as we did in the last section {[}Sec. \eqref{Exterior}{]} in
the case of the exterior region, we decompose $\tilde{f}_{\omega l}^{\mathrm{L}}$
and $\tilde{f}_{\omega l}^{\mathrm{R}}$ {[}related to $f_{\omega lm}^{\mathrm{L}}$
and $f_{\omega lm}^{\mathrm{R}}$ through Eq. \eqref{decom f}{]}
into a time-dependent part and a radial function. In contrast to the
exterior region, in the interior the spacelike and timelike nature
of the coordinates $t$ and $r_{*}$ is interchanged. As a result,
in the case of the interior we use the following decomposition 
\begin{equation}
\tilde{f}_{\omega l}^{\mathrm{L}}\left(r,t\right)=e^{i\omega t}\psi_{\omega l}\left(r\right),\,\,\,\,\,\,\tilde{f}_{\omega l}^{\mathrm{R}}\left(r,t\right)=e^{-i\omega t}\psi_{\omega l}\left(r\right)\label{f L,R decom}
\end{equation}
in terms of a single radial function and two different time-dependent
parts. The radial function $\psi_{\omega l}$ satisfies Eq. \eqref{Radial eq}
and its boundary condition is easily determined from Eqs. \eqref{b.c L mode}
and \eqref{b.c R mode} to be 
\begin{equation}
\psi_{\omega l}\left(r\right)\cong e^{-i\omega r_{*}}\quad,\quad r_{*}\rightarrow-\infty.\label{b.c radial interior}
\end{equation}
Recall that $r_{*}\rightarrow-\infty$ corresponds to $r\rightarrow r_{+}$
and that $t$ and $r_{*}$ are related to $v$ and $u_{\mathrm{int}}$
(in the interior region) through Eq. \eqref{v =000026 u_int}. 

Our main remaining task is to find the asymptotic form of the radial
function $\psi_{\omega l}$ near the inner horizon. Then, using Eqs.
\eqref{G Unruh}, \eqref{G HH}, \eqref{decom f} and \eqref{f L,R decom}
we will get the asymptotic behavior of $\langle\hat{\Phi}^{2}\rangle$
and $\langle\hat{T}_{\mu\nu}\rangle$ near that horizon. 

\section{The radial function in the interior region of the black hole}

We now turn to the discussion of the radial function $\psi_{\omega l}$
in the interior region. Let us remind that this function satisfies
Eq. \eqref{Radial eq} along with the boundary condition given in
Eq. \eqref{b.c radial interior}. Our goal is to find its asymptotic
form near the inner horizon. For that, we divide the interior region
$r_{-}<r<r_{+}$ into three domains and use different methods in order
to get approximations for the radial function in these domains. By
matching the three expressions together we will be able to get the
desired asymptotic form near the inner horizon {[}corresponding to
the boundary condition in Eq. \eqref{b.c radial interior}{]}. As
in Sec. \eqref{Exterior}, we work in terms of $\tilde{l}$ {[}defined
in Eq. \eqref{lt}{]} and take the leading-order behavior in it. We
assume that as in the case of the Boulware state in the external region,
the asymptotic forms of the quantities we consider near the inner
horizon are determined by this large-$\tilde{l}$ limit (even though
the integrand structures of the two mode-sums are different). This
assumption is consistent with numerical results that will be published
elsewhere \cite{Letter}. Explicitly, we assume that for each radial
function $\tilde{l}$ satisfy 
\begin{equation}
\tilde{l}\gg\omega M,\,\frac{r_{+}}{M}.\label{large lt}
\end{equation}

\subsection{Region I ($r\rightarrow r_{+}$)\label{Region I}}

Here we consider the vicinity of the outer horizon. We already looked
at a similar limit in Sec. \eqref{Exterior}, but there we considered
the exterior region and here we look at the interior. The expansion
of the potential given in Eq. \eqref{V exp rp} remains the same,
but since we are inside the BH, $r<r_{+}$, the potential is now negative
(instead of positive) and the asymptotic form of the radial equation
{[}given by Eq. \eqref{Radial eq asym} for the case of the exterior
region{]} becomes 
\begin{equation}
\psi_{\omega l}^{I}{}_{,r_{*}r_{*}}+\left[\omega^{2}+C_{+}\left(r_{+}-r_{-}\right)\exp\left(-2\kappa_{+}r_{+}\right)\exp\left(2\kappa_{+}r_{*}\right)\right]\psi_{\omega l}^{I}=0.\label{Radial eq asym I}
\end{equation}
We are looking for a solution that satisfies the boundary condition
given in Eq. \eqref{b.c radial interior}. This solution, taken to
leading order in $\tilde{l}$, is given by 
\begin{equation}
\psi_{\omega l}^{I}\left(r\right)\cong\tilde{l}^{i\omega/\kappa_{+}}e^{-i\omega r_{+}}\varGamma\left(1-i\omega/\kappa_{+}\right)J_{-i\omega/\kappa_{+}}\left[2\tilde{l}\left(\frac{r_{+}-r}{r_{+}-r_{-}}\right)^{1/2}\right],\label{psi I J}
\end{equation}
where $J_{\alpha}\left(z\right)$ is a Bessel function of the first
kind. Another representation, which will turn out to be interesting
below, takes the form 
\begin{equation}
\psi_{\omega l}^{I}\left(r\right)=p_{1}^{I}F^{I}\left(r\right)+p_{2}^{I}F^{I*}\left(r\right),\label{psi I with K}
\end{equation}
where 
\begin{equation}
F^{I}\left(r\right)=\frac{e^{-i\pi/4}}{\pi}K_{i\omega/\kappa_{+}}\left[-2i\tilde{l}\left(\frac{r_{+}-r}{r_{+}-r_{-}}\right)^{1/2}\right],\label{FI}
\end{equation}
\begin{equation}
p_{1}^{I}=e^{-i\frac{\pi}{4}}\tilde{l}^{i\omega/\kappa_{+}}e^{-i\omega r_{+}}\Gamma\left(1-i\omega/\kappa_{+}\right)e^{-\frac{\pi\omega}{2\kappa_{+}}}\label{pI1}
\end{equation}
and 
\begin{equation}
p_{2}^{I}=e^{i\frac{\pi}{4}}\tilde{l}^{i\omega/\kappa_{+}}e^{-i\omega r_{+}}\Gamma\left(1-i\omega/\kappa_{+}\right)e^{\frac{\pi\omega}{2\kappa_{+}}}.\label{pI2}
\end{equation}

\subsection{Region II (The middle region)\label{Region II}}

We would now like to find the form of the radial function in the region
which is not asymptotically close to the outer (or inner) horizon.
This region is defined as the one in which we can employ the WKB approximation
to a good accuracy in the following way. Defining 
\[
k\left(r\right)\equiv\left(\omega^{2}-V_{l}\left(r\right)\right)^{1/2}
\]
and using the fact that the scale of variation of the potential $V_{l}\left(r\right)$
is given by the mass of the BH $M$, we see from the radial equation
\eqref{Radial eq} that the region (i.e. $r$ values) in which the
(leading order) WKB method yields a good approximation for the radial
function is given by 
\begin{equation}
k\left(r\right)M\gg1.\label{WKB 1}
\end{equation}
Exploiting the fact that we only consider large-$\tilde{l}$ modes
{[}see Eq. \eqref{large lt}{]}, we can write the potential $V_{l}\left(r\right)$
{[}given in Eq. \eqref{Potential}{]} to a leading order in $\tilde{l}$
in the following way: 
\begin{equation}
V_{l}\left(r\right)\cong\left(1-\frac{2M}{r}+\frac{Q^{2}}{r^{2}}\right)\frac{\tilde{l}^{2}}{r^{2}}=-\frac{\tilde{l}^{2}}{r^{4}}\left(r_{+}-r\right)\left(r-r_{-}\right).\label{V large lt}
\end{equation}
Note that the asymptotic form of the potential used in Eqs. \eqref{Radial eq asym}
and \eqref{Radial eq asym I} can be obtained from this form in the
limit $r\rightarrow r_{+}$ (and the large-$\tilde{l}$ limit of $C_{+}$).
Another consequence of Eq. \eqref{large lt} is that we can neglect
$\omega$ in the definition of $k\left(r\right)$, i.e. 
\begin{equation}
k\left(r\right)\cong\left|V_{l}\left(r\right)\right|^{1/2}\cong\frac{\tilde{l}}{r^{2}}\left[\left(r_{+}-r\right)\left(r-r_{-}\right)\right]^{1/2}.\label{k}
\end{equation}
Overall, we can re-write Eq. \eqref{WKB 1} as follows: 
\begin{equation}
\frac{\tilde{l}M}{r^{2}}\left[\left(r_{+}-r\right)\left(r-r_{-}\right)\right]^{1/2}\gg1\label{WKB 2}
\end{equation}
and define region II as the one in which Eq. \eqref{WKB 2} is satisfied. 

Since we are only interested in the leading order in $l$, in this
region we use the leading-order WKB form for the radial function.
We therefore define 
\begin{equation}
\psi_{\omega l}^{II}=k^{-1/2}\left(r\right)\left[a_{+}\exp\left(i\int k\left(r\right)dr_{*}\right)+a_{-}\exp\left(-i\int k\left(r\right)dr_{*}\right)\right],\label{psi II general}
\end{equation}
where the coefficients $a_{+}$ and $a_{-}$ are determined by matching
$\varPsi_{\omega l}^{II}$ and $\varPsi_{\omega l}^{I}$ (see below).
The integral over $k\left(r\right)$ is readily calculated using Eq.
\eqref{k} and we get 
\[
\int k\left(r\right)dr_{*}=-\tilde{l}\arctan\left(\frac{r-M}{\left[\left(r_{+}-r\right)\left(r-r_{-}\right)\right]^{1/2}}\right).
\]
As a result, we have 
\[
\psi_{\omega l}^{II}=\frac{r}{\sqrt{\tilde{l}}}\left[\left(r_{+}-r\right)\left(r-r_{-}\right)\right]^{-1/4}\left\{ a_{+}\exp\left[-i\tilde{l}\arctan\left(\frac{r-M}{\left[\left(r_{+}-r\right)\left(r-r_{-}\right)\right]^{1/2}}\right)\right]\right.
\]
\begin{equation}
\left.+a_{-}\exp\left[i\tilde{l}\arctan\left(\frac{r-M}{\left[\left(r_{+}-r\right)\left(r-r_{-}\right)\right]^{1/2}}\right)\right]\right\} .\label{psi II explicit}
\end{equation}

We can now look at the limit $r_{*}\rightarrow-\infty$ of $\psi_{\omega l}^{II}$
and compare it with the limit $r_{*}\rightarrow\infty$ of $\psi_{\omega l}^{I}$.
The two expressions thus obtained should yield the same result for
$\psi_{\omega l}$ (in this overlapping region) and serve as a way
to obtain $a_{+}$ and $a_{-}$.  Matching the two expressions, we
get (see the appendix for more details) 
\begin{equation}
a_{+}=\left(\frac{\kappa_{+}}{2\pi}\right)^{1/2}i^{\tilde{l}-1}e^{i\frac{\pi}{4}}\tilde{l}^{i\omega/\kappa_{+}}e^{-i\omega r_{+}}\Gamma\left(1-i\omega/\kappa_{+}\right)e^{-\frac{\pi\omega}{2\kappa_{+}}}\label{ap}
\end{equation}
and 
\begin{equation}
a_{-}=\left(\frac{\kappa_{+}}{2\pi}\right)^{1/2}\left(-i\right)^{\tilde{l}}e^{i\frac{\pi}{4}}\tilde{l}^{i\omega/\kappa_{+}}e^{-i\omega r_{+}}\Gamma\left(1-i\omega/\kappa_{+}\right)e^{\frac{\pi\omega}{2\kappa_{+}}}.\label{am}
\end{equation}

An interesting observation is that each exponential term in $\psi_{\omega l}^{II}$
corresponds in the limit $r_{*}\rightarrow-\infty$ (discussed above)
to one of the Bessel functions $K_{\alpha}\left(z\right)$ in the
representation of $\psi_{\omega l}^{I}$ given in Eq. \eqref{psi I with K}
(when taken in the limit $r_{*}\rightarrow\infty$ ). In other words,
in the overlapping region each Bessel function term turns into an
exponent term. 

\subsection{Region III ($r\rightarrow r_{-}$) }

As in region I, we can find from the asymptotic form of the potential
(now near $r_{-}$) the corresponding asymptotic form of the radial
equation. Then, by finding a solution that matches to $\psi_{\omega l}^{II}$
in the overlap between regions II and III, we would obtain $\psi_{\omega l}^{III}$,
which is the asymptotic form of the radial function $\psi_{\omega l}$
near $r_{-}$ and the desired result of this section. 

Expanding the potential near $r_{-}$ {[}as was done in subSec. \eqref{Region I}
and in Sec. \eqref{Exterior} near $r_{+}${]}, we get the following
asymptotic form of the radial equation: 
\begin{equation}
\psi_{\omega l}^{III}{}_{,r_{*}r_{*}}+\left[\omega^{2}-C_{-}\left(r_{+}-r_{-}\right)\exp\left(2\kappa_{-}r_{-}\right)\exp\left(-2\kappa_{-}r_{*}\right)\right]\psi_{\omega l}^{III}=0,\label{Radial eq asym III}
\end{equation}
where 
\[
C_{-}=-\frac{r_{+}-r_{-}}{r_{-}^{4}}\left(l^{2}+l+1-\frac{r_{+}}{r_{-}}\right)
\]
 has the following leading behavior at large $l$ (or $\tilde{l}$):
\[
C_{-}\cong-\frac{2\kappa_{-}}{r_{-}^{2}}\tilde{l}^{2}.
\]
This equation is analogous to Eq. \eqref{Cp large lt} (but note that
$C_{-}$ is negative while $C_{+}$ is positive). 

The general solution (to leading order in $\tilde{l}$) to Eq. \eqref{Radial eq asym III}
can be written as 
\begin{equation}
\psi_{\omega l}^{III}=p_{1}^{III}F^{III}\left(r\right)+p_{2}^{III}F^{III*}\left(r\right),\label{psi 3}
\end{equation}
where 
\begin{equation}
F^{III}\left(r\right)=\frac{e^{-i\pi/4}}{\pi}K_{i\omega/\kappa_{-}}\left[-2i\tilde{l}\left(\frac{r-r_{-}}{r_{+}-r_{-}}\right)^{1/2}\right].\label{F III}
\end{equation}
As stated above, the coefficients $p_{1}^{III}$ and $p_{2}^{III}$
are determined by matching $\psi_{\omega l}^{III}$ in the limit $r_{*}\rightarrow-\infty$
with $\psi_{\omega l}^{II}$ in the limit $r_{*}\rightarrow\infty$.
The result is (see the appendix for more details) 
\begin{equation}
p_{1}^{III}=\frac{r_{-}}{r_{+}}\left(-1\right)^{\tilde{l}}e^{-i\frac{3\pi}{4}}\tilde{l}^{i\omega/\kappa_{+}}e^{-i\omega r_{+}}\Gamma\left(1-i\omega/\kappa_{+}\right)e^{\frac{\pi\omega}{2\kappa_{+}}}\label{pIII1}
\end{equation}
and 
\begin{equation}
p_{2}^{III}=\frac{r_{-}}{r_{+}}\left(-1\right)^{\tilde{l}}e^{-i\frac{\pi}{4}}\tilde{l}^{i\omega/\kappa_{+}}e^{-i\omega r_{+}}\Gamma\left(1-i\omega/\kappa_{+}\right)e^{-\frac{\pi\omega}{2\kappa_{+}}}.\label{pIII2}
\end{equation}

Analogously to the observation made at the end of the last subsection,
here also an exponential term in $\psi_{\omega l}^{II}$ turns in
the overlapping region (between regions II and III) into a Bessel
function term in $\psi_{\omega l}^{III}$. 

An alternative way to write $\psi_{\omega l}^{III}$ in terms of a
different kind of Bessel function is as follows: 
\[
\psi_{\omega l}^{III}=Th\left(r\right)+Rh^{*}\left(r\right),
\]
where 
\[
h\left(r\right)=\tilde{l}^{-i\omega/\kappa_{-}}e^{-i\omega r_{-}}\Gamma\left(1+i\omega/\kappa_{-}\right)J_{i\omega/\kappa_{-}}\left[2\tilde{l}\left(\frac{r-r_{-}}{r_{+}-r_{-}}\right)^{1/2}\right],
\]
\begin{equation}
T=\left(-1\right)^{\tilde{l}-1/2}\left(\frac{r_{-}}{r_{+}}\right)\tilde{l}^{i\omega\left(\kappa_{+}^{-1}+\kappa_{-}^{-1}\right)}e^{-i\omega\left(r_{+}-r_{-}\right)}\frac{\Gamma\left(1-i\omega/\kappa_{+}\right)}{\Gamma\left(1+i\omega/\kappa_{-}\right)}\frac{\sinh\left[\frac{1}{2}\pi\omega\left(\kappa_{+}^{-1}+\kappa_{-}^{-1}\right)\right]}{\sinh\left(\pi\omega/\kappa_{-}\right)},\label{T}
\end{equation}
and 
\begin{equation}
R=\left(-1\right)^{\tilde{l}+1/2}\left(\frac{r_{-}}{r_{+}}\right)\tilde{l}^{i\omega\left(\kappa_{+}^{-1}-\kappa_{-}^{-1}\right)}e^{-i\omega\left(r_{+}+r_{-}\right)}\frac{\Gamma\left(1-i\omega/\kappa_{+}\right)}{\Gamma\left(1-i\omega/\kappa_{-}\right)}\frac{\sinh\left[\frac{1}{2}\pi\omega\left(\kappa_{+}^{-1}-\kappa_{-}^{-1}\right)\right]}{\sinh\left(\pi\omega/\kappa_{-}\right)}.\label{R}
\end{equation}
The advantage of this form is that the function $h\left(r\right)$
satisfies 
\[
h\left(r_{*}\right)\cong e^{-i\omega r_{*}}\quad,\quad r_{*}\rightarrow\infty
\]
when viewed as a function of $r_{*}$. As a result, we have 
\begin{equation}
\psi_{\omega l}^{III}\cong Te^{-i\omega r_{*}}+Re^{i\omega r_{*}}\quad,\quad r_{*}\rightarrow\infty.\label{psi III asym}
\end{equation}
Therefore, if we recall that the ``initial condition'' is given
by Eq. \eqref{b.c radial interior}, we see that we can regard $R$
and $T$ as reflection and transmission coefficients inside the BH,
respectively. 

\subsection{The Wronskian}

A nontrivial check for the above expressions for $\psi_{\omega l}$
in the different regions is the computation of the corresponding Wronskians.
The Wronskian of Eq. \eqref{Radial eq} is conserved, and therefore
we expect the Wronskians calculated with respect to $\psi_{\omega l}^{I}$,
$\psi_{\omega l}^{II}$ and $\psi_{\omega l}^{III}$ to be the same.
In region I, we have {[}using the asymptotic form given in Eq. \eqref{b.c radial interior}{]}
\[
W=2i\mathrm{Im}\left(\psi_{\omega l,r_{*}}^{I}\psi_{\omega l}^{I*}\right)=-2i\omega.
\]
In region II, we can use the general form given in Eq. \eqref{psi II general}
{[}or the explicit form of Eq. \eqref{psi II explicit}{]} and get
\[
W=2i\mathrm{Im}\left(\psi_{\omega l,r_{*}}^{II}\psi_{\omega l}^{II*}\right)=2i\left(\left|a_{+}\right|^{2}-\left|a_{-}\right|^{2}\right).
\]
When substituting $a_{+}$ and $a_{-}$ from Eqs. \eqref{ap} and
\eqref{am}, we indeed obtain the same result as in region I. 

Similarly, in region III we can easily calculate the Wronskian using
the asymptotic form given in Eq. \eqref{psi III asym}. We find 
\[
W=2i\mathrm{Im}\left(\psi_{\omega l,r_{*}}^{III}\psi_{\omega l}^{III*}\right)=-2i\omega\left(\left|T\right|^{2}-\left|R\right|^{2}\right),
\]
and again after the substitution of $T$ and $R$ from Eqs. \eqref{T}
and \eqref{R}, we get the same result as in regions I and II. 

\section{Asymptotic behavior of $\boldsymbol{\langle\hat{\Phi}^{2}\rangle_{ren}}$
and $\boldsymbol{\langle\hat{T}_{\mu\nu}\rangle_{ren}}$ near the
inner horizon}

Now we have all the ingredients that we need in order to calculate
the leading asymptotic behavior of $\langle\hat{\Phi}^{2}\rangle_{ren}$
and $\langle\hat{T}_{\mu\nu}\rangle_{ren}$ near the inner horizon.
We first start from analyzing $\langle\hat{\Phi}^{2}\rangle_{ren}$
and then move to $\langle\hat{T}_{\mu\nu}\rangle_{ren}$. 

\subsection{$\boldsymbol{\langle\hat{\Phi}^{2}\rangle_{ren}}$}

Let us begin by considering the Hartle-Hawking state. In order to
find $\langle\hat{\Phi}^{2}\rangle_{H,ren}$, we use the renormalization
method described in subSec. \eqref{subsec:Reg=000026Ren}, see Eq.
\eqref{phi2 ren}. For that, we first need to find $G_{H}^{\left(1\right)}\left(x,x'\right)$\footnote{Instead of the two-point function $<\hat{\Phi}\left(x\right)\hat{\Phi}\left(x'\right)>$
that appears in Eq. \eqref{phi2 ren}, we here use the Hadamard function
$G_{HH}^{\left(1\right)}\left(x,x'\right)$. Therefore, we need to
include an extra factor of $1/2$ in front of $G_{HH}^{\left(1\right)}\left(x,x'\right)$
in Eq. \eqref{phi2 ren}. }, conveniently expressed in Eq. \eqref{G HH} in terms of the inner
Eddington-Finkelstein modes $f_{\omega lm}^{\mathrm{L}}\left(x\right)$
and $f_{\omega lm}^{\mathrm{R}}\left(x\right)$. As mentioned in subSec.
\eqref{subsec:Reg=000026Ren}, we take the separation between the
two points $x$ and $x'$ to be in the $t$ direction. Specifically,
we choose $x=\left(t,r,\theta,\varphi\right)$ and $x'=\left(t+\varepsilon,r,\theta,\varphi\right)$.
Then, substituting $\psi_{\omega l}^{III}\left(r\right)$ given by
Eq. \eqref{psi 3} into Eq. \eqref{f L,R decom}, and the resulting
$\tilde{f}_{\omega l}^{\mathrm{L}}\left(t,r\right)$ and $\tilde{f}_{\omega l}^{\mathrm{R}}\left(t,r\right)$
into Eq. \eqref{decom f}, we obtain the inner Eddington-Finkelstein
modes $f_{\omega lm}^{\mathrm{L}}\left(x\right)$ and $f_{\omega lm}^{\mathrm{R}}\left(x\right)$
near the inner horizon (i.e. in region III). Next, we substitute these
modes {[}along with the expression for $\rho_{\omega l}^{\mathrm{up}}$
we found in Eq. \eqref{Rho}{]} into Eq. \eqref{G HH} and get 
\[
G_{H}^{\left(1\right)}\left(x,x'\right)\cong\frac{8\pi}{\kappa_{-}}\intop_{0}^{\infty}d\omega\cos\left(\omega\varepsilon\right)\sum_{l,m}\left|C_{lm}\right|^{2}\left|F^{III}\left(r\right)\right|^{2},\,\,\,r\rightarrow r_{-},
\]
where $F^{III}\left(r\right)$ is given by Eq. \eqref{F III} and
$C_{lm}$ by Eq. \eqref{Ct}. As in Sec. \eqref{Exterior}, we take
advantage of the fact that the leading contribution comes from large-$l$
values and switch the summation over $l$ with an integral. After
we substitute for $C_{lm}$ using Eq. \eqref{Ct} and perform the
summation over $m$, we find 
\begin{equation}
G_{H}^{\left(1\right)}\left(x,x'\right)\cong\frac{2}{\pi\left(r_{+}-r_{-}\right)}\intop_{0}^{\infty}d\omega\cos\left(\omega\varepsilon\right)\intop_{0}^{\infty}\left|F^{III}\left(r\right)\right|^{2}\tilde{l}d\tilde{l},\,\,\,r\rightarrow r_{-}.\label{G HH calc}
\end{equation}
The integral over $\tilde{l}$ in this expression is divergent. In
order to find its correct value, we proceed as follows.  We start
by writing it in the form 
\[
I_{0}\equiv\intop_{0}^{\infty}\left|F^{III}\left(r\right)\right|^{2}\tilde{l}d\tilde{l}=\pi^{-2}\intop_{0}^{\infty}K_{i\omega/\kappa_{-}}\left[-2i\tilde{l}\left(\frac{r-r_{-}}{r_{+}-r_{-}}\right)^{1/2}\right]K_{i\omega/\kappa_{-}}\left[2i\tilde{l}\left(\frac{r-r_{-}}{r_{+}-r_{-}}\right)^{1/2}\right]\tilde{l}d\tilde{l}
\]
\[
\equiv\pi^{-2}\intop_{0}^{\infty}K_{\nu}\left(-\mu\tilde{l}\right)K_{\nu}\left(\mu\tilde{l}\right)\tilde{l}d\tilde{l},
\]
where we used the fact that $\left[K_{ic}\left(id\right)\right]^{*}=K_{ic}\left(-id\right)$
for $c,d\in\mathbb{R}$, and defined 
\[
\mu\equiv2i\left(\frac{r-r_{-}}{r_{+}-r_{-}}\right)^{1/2},\quad\nu\equiv i\omega/\kappa_{-}.
\]
Next, we regulate this integral by adding a small, real and positive
$\delta$ to the arguments of the Bessel functions, and get 
\[
I_{0}=\pi^{-2}\intop_{0}^{\infty}K_{\nu}\left(-\mu\tilde{l}\right)K_{\nu}\left(\mu\tilde{l}\right)\tilde{l}d\tilde{l}\rightarrow I_{\delta}\equiv\pi^{-2}\intop_{0}^{\infty}K_{\nu}\left[\left(-\mu+\delta\right)\tilde{l}\right]K_{\nu}\left[\left(\mu+\delta\right)\tilde{l}\right]\tilde{l}d\tilde{l}.
\]
We can now use the formula 
\[
\intop_{0}^{\infty}K_{\nu}\left(ax\right)K_{\nu}\left(bx\right)xdx=\frac{\pi\left(ab\right)^{-\nu}\left(a^{2\nu}-b^{2\nu}\right)}{2\sin\left(\pi\nu\right)\left(a^{2}-b^{2}\right)},
\]
valid for 
\[
\left|\mathrm{Re}\left(\nu\right)\right|<1,\quad\mathrm{Re}\left(a+b\right)>0,
\]
and obtain 

\[
I_{\delta}=\frac{\left(-\mu^{2}+\delta^{2}\right)^{-\nu}\left[\left(-\mu+\delta\right)^{2\nu}-\left(\mu+\delta\right)^{2\nu}\right]}{2\pi\sin\left(\pi\nu\right)\left[\left(-\mu+\delta\right)^{2}-\left(\mu+\delta\right)^{2}\right]}.
\]
Expanding $I_{\delta}$ in powers of $\delta$ and substituting the
expressions for $\mu$ and $\nu$, we have
\[
I_{\delta}=\frac{1}{8\pi\delta}\left(\frac{r-r_{-}}{r_{+}-r_{-}}\right)^{-1/2}-\frac{\omega}{8\pi\kappa_{-}}\left(\frac{r-r_{-}}{r_{+}-r_{-}}\right)^{-1}\coth\left(\frac{\pi\omega}{\kappa_{-}}\right)+\mathcal{O}\left(\delta\right).
\]
We see that the divergent (as $\delta\rightarrow0$) term in $I_{\delta}$
is independent of $\omega$ and thus does not contribute to $G_{H}^{\left(1\right)}\left(x,x'\right)$
when substituted in Eq. \eqref{G HH calc} (this term is a ``blind
spot'' in the language of \cite{Adam T 2}). As a result, we can
remove this term and then take the limit $\delta\rightarrow0$. The
resulting regularized integral over $\tilde{l}$ is 
\[
\left[I_{0}\right]_{reg}=-\frac{\omega}{8\pi\kappa_{-}}\left(\frac{r-r_{-}}{r_{+}-r_{-}}\right)^{-1}\coth\left(\frac{\pi\omega}{\kappa_{-}}\right),
\]
and after substituting it to Eq. \eqref{G HH calc}, we get 
\begin{equation}
G_{H}^{\left(1\right)}\left(x,x'\right)\cong\frac{2}{\pi\left(r_{+}-r_{-}\right)}\intop_{0}^{\infty}d\omega\cos\left(\omega\varepsilon\right)\left[I_{0}\right]_{reg}=-\frac{1}{4\pi^{2}\kappa_{-}\left(r-r_{-}\right)}\intop_{0}^{\infty}\omega\coth\left(\frac{\pi\omega}{\kappa_{-}}\right)\cos\left(\omega\varepsilon\right)d\omega,\,\,\,r\rightarrow r_{-}.\label{G HH mode-sum}
\end{equation}

We can now move on to consider the second term in the right hand side
of Eq. \eqref{phi2 ren}, the DeWitt-Schwinger counterterm $G_{DS}\left(x,x'\right)$.
We have a massless ($m=0$), minimally-coupled ($\xi=0$) scalar field
in Reissner-Nordström spacetime ($R=0$). Therefore, substituting
$m=\xi=R=0$ in Eq. \eqref{G DS}, we get 
\[
G_{DS}\left(x,x'\right)=\frac{1}{8\pi^{2}\sigma}+\frac{1}{96\pi^{2}}R_{\mu\nu}\frac{\sigma^{;\mu}\sigma^{;\nu}}{\sigma}.
\]
Now, recall that $\sigma$ is the biscalar of geodetic separation
and that we took the points $x$ and $x'$ to be separated along the
$t$ direction: $x=\left(t,r,\theta,\varphi\right)$ and $x'=\left(t+\varepsilon,r,\theta,\varphi\right)$.
Then, we can look at the (shortest) geodesic that connects $x$ and
$x'$ and expand it in powers of $\varepsilon$. For a general metric
function $f\left(r\right)$ {[}i.e. $\left(1-2M/r+Q^{2}/r^{2}\right)\rightarrow f\left(r\right)${]},
we get the following expansion of $G_{DS}\left(x,x'\right)$ in powers
of $\varepsilon$: 
\[
G_{DS}\left(x,x'\right)=-\frac{1}{4\pi^{2}f\left(r\right)}\varepsilon^{-2}+\frac{f'\left(r\right)^{2}}{192\pi^{2}f\left(r\right)}-\frac{1}{48\pi^{2}f\left(r\right)}R_{00}+\mathcal{O}\left(\varepsilon^{2}\right),
\]
and for the Reissner-Nordström metric, $f\left(r\right)=\left(1-2M/r+Q^{2}/r^{2}\right)$,
we obtain near $r=r_{-}${[}henceforth, we remove the $\mathcal{O}\left(\varepsilon^{2}\right)$
terms{]} 
\[
G_{DS}\left(x,x'\right)\cong\frac{1}{8\pi^{2}\kappa_{-}\left(r-r_{-}\right)}\varepsilon^{-2}-\frac{\kappa_{-}}{96\pi^{2}\left(r-r_{-}\right)},\,\,\,r\rightarrow r_{-}.
\]
Using 
\[
\varepsilon^{-2}=-\intop_{0}^{\infty}\omega\cos\left(\omega\varepsilon\right)d\omega,
\]
we can also write 
\begin{equation}
G_{DS}\left(x,x'\right)\cong-\frac{1}{8\pi^{2}\kappa_{-}\left(r-r_{-}\right)}\intop_{0}^{\infty}\omega\cos\left(\omega\varepsilon\right)d\omega-\frac{\kappa_{-}}{96\pi^{2}\left(r-r_{-}\right)},\,\,\,r\rightarrow r_{-}.\label{G DS mode-sum}
\end{equation}

Now, we can finally find the leading behavior of $\langle\hat{\Phi}^{2}\rangle_{H,ren}$
near $r=r_{-}$. Substituting Eqs. \eqref{G HH mode-sum} and \eqref{G DS mode-sum}
into \eqref{phi2 ren}, we get 
\[
\left\langle \hat{\Phi}^{2}\left(x\right)\right\rangle _{H,ren}=\lim_{x'\rightarrow x}\left[\left\langle \hat{\Phi}\left(x\right)\hat{\Phi}\left(x'\right)\right\rangle _{H}-G_{DS}\left(x,x'\right)\right]=\lim_{x'\rightarrow x}\left[\frac{1}{2}G_{H}^{\left(1\right)}\left(x,x'\right)-G_{DS}\left(x,x'\right)\right]
\]
\begin{equation}
\cong-\frac{1}{8\pi^{2}\kappa_{-}\left(r-r_{-}\right)}\intop_{0}^{\infty}\omega\left[\coth\left(\frac{\pi\omega}{\kappa_{-}}\right)-1\right]d\omega+\frac{\kappa_{-}}{96\pi^{2}\left(r-r_{-}\right)}=0,\,\,\,r\rightarrow r_{-}.\label{phi2 H ren}
\end{equation}
Note that in the second line the limit $\varepsilon\rightarrow0$
was taken before the integration was carried out. Of course, we could
have performed the integration first (using the Abel-summation integral,
see \cite{Adam phi2 t}) and only then take the $\varepsilon\rightarrow0$
limit and get the same result. We obtained that the coefficient of
the expected leading divergence {[}$\propto\left(r-r_{-}\right)^{-1}${]}
of $\langle\hat{\Phi}^{2}\rangle_{H,ren}$ near the inner horizon
\emph{vanishes}. As a result, according to this analysis $\langle\hat{\Phi}^{2}\rangle_{H,ren}$
may have a weaker divergence, such as $\propto\log\left(r-r_{-}\right)$,
or it may be regular. This is consistent with new numerical results
\cite{Letter} showing that $\langle\hat{\Phi}^{2}\rangle_{H,ren}$
approaches a finite value at the inner horizon and is therefore regular. 

As for the Unruh state, we obtain from subtracting Eqs. \eqref{G HH}
and \eqref{G Unruh} that
\[
G_{H}^{\left(1\right)}\left(x,x'\right)-G_{U}^{\left(1\right)}\left(x,x'\right)=\intop_{0}^{\infty}d\omega\sum_{l,m}\left|\tau_{\omega l}^{\mathrm{up}}\right|^{2}\left[\coth\left(\frac{\pi\omega}{\kappa_{+}}\right)-1\right]\left\{ f_{\omega lm}^{\mathrm{R}}\left(x\right),f_{\omega lm}^{\mathrm{R}*}\left(x'\right)\right\} .
\]
Then, from Eq. \eqref{phi2 ren} we get 
\[
\left\langle \hat{\Phi}^{2}\left(x\right)\right\rangle _{H,ren}-\left\langle \hat{\Phi}^{2}\left(x\right)\right\rangle _{U,ren}=\intop_{0}^{\infty}d\omega\sum_{l,m}\left|\tau_{\omega l}^{\mathrm{up}}\right|^{2}\left[\coth\left(\frac{\pi\omega}{\kappa_{+}}\right)-1\right]\left|f_{\omega lm}^{\mathrm{R}}\left(x\right)\right|^{2}.
\]
This quantity does not contribute at the leading order, since its
dominant part comes from small $\omega$ and large $l$ values, and
is highly suppressed by $\tau_{\omega l}^{\mathrm{up}}$ in this domain.
Therefore, the leading divergence of $\langle\hat{\Phi}^{2}\rangle_{U,ren}$
will be the same as that of $\langle\hat{\Phi}^{2}\rangle_{H,ren}$,
which is vanishing. This again matches the numerical results \cite{Letter}.

\subsection{$\boldsymbol{\langle\hat{T}_{\mu\nu}\rangle_{ren}}$}

The same argument that appears in the end of the previous subsection
for $\hat{\Phi}^{2}$ applies to $\hat{T}_{\mu\nu}$ as well, hence
the leading divergence of the stress-energy tensor should be the same
for both the Unruh and Hartle-Hawking states. As in the previous subsection,
we choose to calculate the expectation value in the Hartle-Hawking
state (because the mode-sum expressions are less complicated). We
follow the renormalization prescription mentioned in Sec. \eqref{subsec:Reg=000026Ren}. 

Instead of calculating each of the components of $\langle\hat{T}_{\mu\nu}\rangle_{H,ren}$
independently, we can use the conservation of stress-energy and Eq.
\eqref{Trace anomaly} in order to relate the various components,
ending up with only one independent component. To see this, note that
we found that the ``leading divergence'' of $\langle\hat{\Phi}^{2}\rangle_{H,ren}$
near the inner horizon is $\propto\left(r-r_{-}\right)^{-1}$ {[}see,
for example, Eq. \eqref{G HH mode-sum} or \eqref{phi2 H ren}{]}
and that the corresponding coefficient vanishes. As a result, we expect
the leading divergence of $\langle\hat{T}_{\mu\nu}\rangle_{H,ren}$
to be $\propto\left(r-r_{-}\right)^{-2}$ with a potentially nonvanishing
coefficient {[}similarly to $\langle\hat{T}_{\mu\nu}\rangle_{B,ren}$
in the exterior region near $r=r_{+}$, see Eq. \eqref{T B ren outside}{]}.
Recalling that we have a spherical symmetry, we can therefore write
\begin{equation}
\left\langle \hat{T}_{\nu}^{\mu}\right\rangle _{H,ren}\cong\left(r-r_{-}\right)^{-2}\left(\begin{array}{cccc}
c_{t} & 0 & 0 & 0\\
0 & c_{r} & 0 & 0\\
0 & 0 & c_{\theta} & 0\\
0 & 0 & 0 & c_{\theta}
\end{array}\right),\,\,\,r\rightarrow r_{-}.\label{T H ren asym rm}
\end{equation}
In order to use the conservation of energy and momentum, we quote
the following formula: 
\[
A_{\nu;\mu}^{\mu}=\frac{1}{\sqrt{-g}}\left(\sqrt{-g}A_{\nu}^{\mu}\right)_{,\mu}-\frac{1}{2}g_{\mu\sigma,\nu}A^{\mu\sigma},
\]
valid for a general rank-2 symmetric tensor $A_{\nu}^{\mu}$. Applying
it to the conservation equation 
\[
\left\langle \hat{T}_{\nu;\mu}^{\mu}\right\rangle _{H,ren}=0
\]
and choosing $\nu=r$, we get at leading order 
\[
0=\left\langle \hat{T}_{r;\mu}^{\mu}\right\rangle _{H,ren}\cong\left\langle \hat{T}_{r,r}^{r}\right\rangle _{H,ren}-\frac{1}{2}g_{tt,r}\left\langle \hat{T}^{tt}\right\rangle _{H,ren}-\frac{1}{2}g_{rr,r}\left\langle \hat{T}^{rr}\right\rangle _{H,ren}\cong-\frac{1}{2}\left(r-r_{-}\right)^{-3}\left(c_{t}+3c_{r}\right),\,\,\,r\rightarrow r_{-},
\]
hence 
\begin{equation}
c_{t}+3c_{r}=0.\label{c1}
\end{equation}

Next, we would like to use Eq. \eqref{Trace anomaly}. Since $T_{\mathrm{anomaly}}$
is a local, geometric quantity built from curvature scalars {[}see
Eq. \eqref{T_anomaly}{]}, it is regular at the inner horizon. Moreover,
since the divergence (if any) of $\langle\hat{\Phi}^{2}\rangle_{H,ren}$
at $r=r_{-}$ is weaker than $\propto\left(r-r_{-}\right)^{-1}$,
the divergence of $\square\langle\hat{\Phi}^{2}\rangle_{H,ren}$ is
weaker than $\propto\left(r-r_{-}\right)^{-2}$ \footnote{For example, if $<\hat{\Phi}^{2}>_{H,ren}\propto\log\left(r-r_{-}\right)$
than $\square<\hat{\Phi}^{2}>_{H,ren}$ is regular at $r=r_{-}$.}. Therefore, according to Eq. \eqref{Trace anomaly}, the trace of
the stress-energy tensor vanishes at leading order {[}$\propto\left(r-r_{-}\right)^{-2}${]}
and we have from Eq. \eqref{T H ren asym rm}: 
\begin{equation}
c_{t}+c_{r}+2c_{\theta}=0.\label{c2}
\end{equation}
Combining Eqs. \eqref{c1} and \eqref{c2}, we obtain 
\begin{equation}
c_{r}=c_{\theta}=-\frac{1}{3}c_{t}.\label{c3}
\end{equation}
As mentioned above, we see that there is only one independent component
of $\langle\hat{T}_{\mu\nu}\rangle_{H,ren}$ at leading order. We
choose to calculate $\langle\hat{T}_{tt}\rangle_{H,ren}$. Now, since
the trace $\langle\hat{T}_{\mu}^{\mu}\rangle_{H,ren}$ does not contribute
at leading order, it is enough to calculate $\langle\{\hat{\Phi}_{,t}\left(x\right)\hat{\Phi}_{,t}\left(x'\right)\}\rangle_{H}$
instead of the whole expression for $\langle\hat{T}_{tt}\left(x,x'\right)\rangle_{H}$
\footnote{Notice that, as discussed in subSec. \eqref{Point-Splitting}, all
the derivatives in the expression for the stress-energy tensor are
taken at the same point $x$, and the bi-vector of parallel transport
is absent (it appears in the counterterm instead).}. As before, we take the splitting to be in the $t$-direction and
choose $x=\left(t,r,\theta,\varphi\right)$ and $x'=\left(t+\varepsilon,r,\theta,\varphi\right)$.
It is easy to see that the mode-sum expression for this quantity is
the same as that of $G_{H}^{\left(1\right)}\left(x,x'\right)$, given
in Eq. \eqref{G HH mode-sum}, up to an extra factor of $\omega^{2}$
in the integrand. Thus, we have 
\begin{equation}
\left\langle \hat{T}_{tt}\left(x,x'\right)\right\rangle _{H}\cong\frac{1}{2}\left\langle \left\{ \hat{\Phi}_{,t}\left(x\right),\hat{\Phi}_{,t}\left(x'\right)\right\} \right\rangle _{H}\cong-\frac{1}{8\pi^{2}\kappa_{-}\left(r-r_{-}\right)}\intop_{0}^{\infty}\omega^{3}\coth\left(\frac{\pi\omega}{\kappa_{-}}\right)\cos\left(\omega\varepsilon\right)d\omega,\,\,\,r\rightarrow r_{-}.\label{T split asym}
\end{equation}

In \cite{Adam T 2}, the general form of the counterterm $C_{\mu\nu}^{DS}$
in terms of Christensen's original one \cite{Christensen 78} was
found, and its expansion in powers of $\varepsilon$ (the $t$-splitting
parameter) was explicitly obtained for a massless, minimally coupled
scalar field in Schwarzschild spacetime. The extension of this result
to Reissner-Nordström spacetime is given by\footnote{I thank A. Levi for providing this counterterm for me.}
\begin{equation}
C_{\mu\nu}^{DS}\left(x,x'\right)=\intop_{0}^{\infty}\left(a_{\mu\nu}\omega^{3}+b_{\mu\nu}\omega+c_{\mu\nu}\ln\left(\omega\right)+d_{\mu\nu}\frac{1}{\omega+\mu e^{-\gamma}}\right)\cos\left(\omega\varepsilon\right)d\omega+e_{\mu\nu},\label{C DS RN}
\end{equation}
where the coefficients $b_{tt}$, $c_{tt}$ and $d_{tt}$ do not contribute
at leading order and 
\[
a_{tt}\cong-\frac{1}{8\pi^{2}\kappa_{-}\left(r-r_{-}\right)},\,\,\,r\rightarrow r_{-},
\]
\[
e_{tt}\cong-\frac{\kappa_{-}^{3}}{960\pi^{2}\left(r-r_{-}\right)},\,\,\,r\rightarrow r_{-}.
\]

Substituting the expressions we obtained for the counterterm and the
(split) stress-energy tensor {[}given by Eqs. \eqref{T split asym}
and \eqref{C DS RN}{]} into Eq. \eqref{T ren} and taking the limit
$\varepsilon\rightarrow0$, we get 
\[
\left\langle \hat{T}_{tt}\right\rangle _{H,ren}\cong-\frac{1}{8\pi^{2}\kappa_{-}\left(r-r_{-}\right)}\intop_{0}^{\infty}\left[\coth\left(\frac{\pi\omega}{\kappa_{-}}\right)-1\right]\omega^{3}d\omega+\frac{\kappa_{-}^{3}}{960\pi^{2}\left(r-r_{-}\right)}=0,\,\,\,r\rightarrow r_{-}.
\]
Using Eq. \eqref{c3}, we find that the coefficient of the expected
leading divergence of all the components of $\langle\hat{T}_{\mu\nu}\rangle_{H,ren}$
near the inner horizon \emph{vanishes}. As mentioned above, the same
applies to $\langle\hat{T}_{\mu\nu}\rangle_{U,ren}$. 

\section{Discussion}

In this work we considered a massless, minimally coupled quantum scalar
field on a RN black hole background, and studied the asymptotic behavior
of $\langle\hat{\Phi}^{2}\rangle_{ren}$ and $\langle\hat{T}_{\mu\nu}\rangle_{ren}$
near the inner and outer horizons in quantum states in which they
are expected to diverge. Our strategy was to analyze the modes of
the scalar field near the horizons, where analytic expressions can
be obtained. Then, using expressions for these expectation values
as mode-sums of these modes, we obtained their leading asymptotic
behavior near the horizons. In this calculation we made the assumption
that this asymptotic behavior is determined by the large-$l$ limit
of the modes, and we found agreement with \cite{Anderson,Anderson phi2,Letter}. 

In section III, we used this analytical approach to obtain the known
divergence of $\langle\hat{\Phi}^{2}\rangle_{ren}$ and $\langle\hat{T}_{\mu\nu}\rangle_{ren}$,
evaluated in Boulware state, at the event horizon. We derived new
and explicit analytic expressions for the asymptotic forms of these
expectation values, which correspond to the numerical results obtained
in Refs. \cite{Anderson,Anderson phi2}. 

Then, in the rest of the paper, we applied our approach to the calculation
of the asymptotic behavior of $\langle\hat{\Phi}^{2}\rangle_{ren}$
and $\langle\hat{T}_{\mu\nu}\rangle_{ren}$, evaluated in Unruh and
Hartle-Hawking states, near the inner (Cauchy) horizon. We found that
the coefficient of this ``leading'' divergence vanishes, and therefore
the divergence, if it occurs, is weaker than expected a priori. These
a priori expectations may originate from various different directions.
First, by examining the expressions for the counterterms or the mode-sums
near the inner horizon {[}see, for example, Eq. \eqref{G DS mode-sum}{]},
we might expect these strong divergences. For example, in the case
of $\langle\hat{\Phi}^{2}\rangle_{ren}$, this yields a $\propto\left(r-r_{-}\right)^{-1}$
divergence. Second, these strong divergences exactly correspond to
the ones found in section III in the case of the Boulware state and
the outer (event) horizon. One may expect that this is a general behavior
of this kinds of expectation values near horizons. Finally, the same
strong divergences were obtained in Ref. \cite{Huang} for a conformally
(rather than minimally) coupled scalar field, but under a certain
approximation that was carried out. Our goal was thus to calculate
the leading asymptotic forms of $\langle\hat{\Phi}^{2}\rangle_{ren}$
and $\langle\hat{T}_{\mu\nu}\rangle_{ren}$ near the inner horizon
without recourse to the approximation methods used before and for
a minimally-coupled scalar field. 

In the case of the Unruh state (describing an evaporating BH), it
was shown in Ref. \cite{Hiscock} that the RSET has to diverge on
at least one of the inner horizons (in the RN case). Therefore, our
findings show that this divergence is weaker than might be expected
a priori. This, in turn, opens the door for a scenario in which the
resulting modification of the metric is finite. 

In order to obtain the exact asymptotic form near the horizon (and
not only the leading-order one), numerical calculation and mode-sum
of the modes should be employed. This study will be published in a
subsequent paper and its results match the ones obtained in this note
\cite{Letter}. 

It will be interesting to apply our analytical approach to the calculation
of the asymptotic form of the RSET of a conformally coupled scalar
field, and test whether the approximate results of \cite{Huang} are
valid near the inner horizon. Furthermore, an extension to a more
realistic model, such as to a Kerr black hole background or a quantum
electromagnetic field instead of a scalar one, will add an important
contribution to the picture. 

\section*{ACKNOWLEDGMENTS}

I would like to thank Amos Ori for his guidance throughout the execution
of this work. This research was supported by the Israel Science Foundation
under Grant No. 1696/15 and by the I-CORE Program of the Planning
and Budgeting Committee. 

\section*{APPENDIX: THE RADIAL FUNCTION IN THE INTERNAL REGION OF THE BLACK
HOLE}

In this appendix we derive the expressions \eqref{ap}, \eqref{am},
\eqref{pIII1} and \eqref{pIII2} for the coefficients in the definitions
of $\psi_{\omega l}^{II}$ and $\psi_{\omega l}^{III}$. The strategy
is to take the $r_{*}\rightarrow\pm\infty$ limits of $\psi_{\omega l}^{II}$
and compare it with the limits of $\psi_{\omega l}^{I}$ and $\psi_{\omega l}^{III}$
corresponding to the overlapping regions with $\psi_{\omega l}^{II}$.
We begin with the expression for $\psi_{\omega l}^{II}$ given in
Eq. \eqref{psi II explicit}, which we reproduce here for convenience,
\[
\psi_{\omega l}^{II}=\frac{r}{\sqrt{\tilde{l}}}\left[\left(r_{+}-r\right)\left(r-r_{-}\right)\right]^{-1/4}\left\{ a_{+}\exp\left[-i\tilde{l}\arctan\left(\frac{r-M}{\left[\left(r_{+}-r\right)\left(r-r_{-}\right)\right]^{1/2}}\right)\right]\right.
\]
\[
\left.+a_{-}\exp\left[i\tilde{l}\arctan\left(\frac{r-M}{\left[\left(r_{+}-r\right)\left(r-r_{-}\right)\right]^{1/2}}\right)\right]\right\} .
\]
In order to obtain $\psi_{\omega l}^{II}$ at the limits $r\rightarrow r_{\pm}$,
we first consider the exponents at these limits, 
\[
\arctan\left(\frac{r-M}{\sqrt{\left(r_{+}-r\right)\left(r-r_{-}\right)}}\right)=\left\{ \begin{array}{c}
\frac{\pi}{2}-2\left(\frac{r_{+}-r}{r_{+}-r_{-}}\right)^{1/2}+\mathcal{O}\left[\left(r_{+}-r\right)^{3/2}\right]\,,\,\,\,r\rightarrow r_{+}\\
\\
-\frac{\pi}{2}+2\left(\frac{r-r_{-}}{r_{+}-r_{-}}\right)^{1/2}+\mathcal{O}\left[\left(r-r_{-}\right)^{3/2}\right]\,,\,\,\,r\rightarrow r_{-}
\end{array}\right.,
\]
from which we get 
\[
\psi_{\omega l}^{II}\cong\frac{r_{+}}{\sqrt{\tilde{l}}}\left[\left(r_{+}-r\right)\left(r_{+}-r_{-}\right)\right]^{-1/4}\left\{ a_{+}\left(-i\right)^{\tilde{l}}\exp\left[2i\tilde{l}\left(\frac{r_{+}-r}{r_{+}-r_{-}}\right)^{1/2}\right]+\right.
\]
\begin{equation}
\left.+a_{-}i^{\tilde{l}}\exp\left[-2i\tilde{l}\left(\frac{r_{+}-r}{r_{+}-r_{-}}\right)^{1/2}\right]\right\} \,,\,\,\,r\rightarrow r_{+}\label{app psi II asym rp}
\end{equation}
and 
\[
\psi_{\omega l}^{II}\cong\frac{r_{-}}{\sqrt{\tilde{l}}}\left[\left(r_{+}-r_{-}\right)\left(r-r_{-}\right)\right]^{-1/4}\left\{ a_{+}i^{\tilde{l}}\exp\left[-2i\tilde{l}\left(\frac{r-r_{-}}{r_{+}-r_{-}}\right)^{1/2}\right]+\right.
\]
\begin{equation}
\left.+a_{-}\left(-i\right)^{\tilde{l}}\exp\left[2i\tilde{l}\left(\frac{r-r_{-}}{r_{+}-r_{-}}\right)^{1/2}\right]\right\} \,,\,\,\,r\rightarrow r_{-}.\label{app psi II asym rm}
\end{equation}

Next, to get $\psi_{\omega l}^{I}$ and $\psi_{\omega l}^{III}$ in
the overlapping regions, we take the $r\rightarrow-\infty$ limit
of $\psi_{\omega l}^{I}$ and the $r\rightarrow\infty$ limit of $\psi_{\omega l}^{III}$.
Using the following asymptotic form of the Bessel K function, 
\[
K_{c}\left(y\right)\cong\sqrt{\frac{\pi}{2y}}\exp\left(-y\right)\,,\,\,\,y\rightarrow\infty
\]
valid for any complex number $c$, we can find the desired asymptotic
forms of $F^{I}\left(r\right)$ and $F^{III}\left(r\right)$ given
in Eqs. \eqref{FI} and \eqref{F III}, 
\[
F^{I}\left(r\right)\cong\frac{1}{2\sqrt{\pi\tilde{l}}}\left(\frac{r_{+}-r}{r_{+}-r_{-}}\right)^{-1/4}\exp\left[2i\tilde{l}\left(\frac{r_{+}-r}{r_{+}-r_{-}}\right)^{1/2}\right]\,,\,\,\,r\rightarrow-\infty,
\]
\[
F^{III}\left(r\right)\cong\frac{1}{2\sqrt{\pi\tilde{l}}}\left(\frac{r-r_{-}}{r_{+}-r_{-}}\right)^{-1/4}\exp\left[2i\tilde{l}\left(\frac{r-r_{-}}{r_{+}-r_{-}}\right)^{1/2}\right]\,,\,\,\,r\rightarrow\infty.
\]
Substituting these expressions in \eqref{psi I with K} and \eqref{psi 3},
we obtain 
\begin{equation}
\psi_{\omega l}^{I}\left(r\right)\cong\frac{1}{2\sqrt{\pi\tilde{l}}}\left(\frac{r_{+}-r}{r_{+}-r_{-}}\right)^{-1/4}\left\{ p_{1}^{I}\exp\left[2i\tilde{l}\left(\frac{r_{+}-r}{r_{+}-r_{-}}\right)^{1/2}\right]+p_{2}^{I}\exp\left[-2i\tilde{l}\left(\frac{r_{+}-r}{r_{+}-r_{-}}\right)^{1/2}\right]\right\} \,,\,\,\,r\rightarrow-\infty\label{app psi I asym}
\end{equation}
and 
\begin{equation}
\psi_{\omega l}^{III}\cong\frac{1}{2\sqrt{\pi\tilde{l}}}\left(\frac{r-r_{-}}{r_{+}-r_{-}}\right)^{-1/4}\left\{ p_{1}^{III}\exp\left[2i\tilde{l}\left(\frac{r-r_{-}}{r_{+}-r_{-}}\right)^{1/2}\right]+p_{2}^{III}\exp\left[-2i\tilde{l}\left(\frac{r-r_{-}}{r_{+}-r_{-}}\right)^{1/2}\right]\right\} \,,\,\,\,r\rightarrow\infty.\label{app psi III asym}
\end{equation}

We can now compare \eqref{app psi II asym rp} with \eqref{app psi I asym},
and \eqref{app psi II asym rm} with \eqref{app psi III asym}. We
find 
\[
a_{+}=\left(\frac{\kappa_{+}}{2\pi}\right)^{1/2}i^{\tilde{l}}p_{1}^{I}=\left(\frac{\kappa_{+}}{2\pi}\right)^{1/2}i^{\tilde{l}-1}e^{i\frac{\pi}{4}}\tilde{l}^{i\omega/\kappa_{+}}e^{-i\omega r_{+}}\Gamma\left(1-i\omega/\kappa_{+}\right)e^{-\frac{\pi\omega}{2\kappa_{+}}},
\]
\[
a_{-}=\left(\frac{\kappa_{+}}{2\pi}\right)^{1/2}\left(-i\right)^{\tilde{l}}p_{2}^{I}=\left(\frac{\kappa_{+}}{2\pi}\right)^{1/2}\left(-i\right)^{\tilde{l}}e^{i\frac{\pi}{4}}\tilde{l}^{i\omega/\kappa_{+}}e^{-i\omega r_{+}}\Gamma\left(1-i\omega/\kappa_{+}\right)e^{\frac{\pi\omega}{2\kappa_{+}}},
\]
\[
p_{1}^{III}=\left(\frac{\kappa_{-}}{2\pi}\right)^{-1/2}\left(-i\right)^{\tilde{l}}a_{-}=\frac{r_{-}}{r_{+}}\left(-1\right)^{\tilde{l}}e^{-i\frac{3\pi}{4}}\tilde{l}^{i\omega/\kappa_{+}}e^{-i\omega r_{+}}\Gamma\left(1-i\omega/\kappa_{+}\right)e^{\frac{\pi\omega}{2\kappa_{+}}},
\]
\[
p_{2}^{III}=\left(\frac{\kappa_{-}}{2\pi}\right)^{-1/2}i^{\tilde{l}}a_{+}=\frac{r_{-}}{r_{+}}\left(-1\right)^{\tilde{l}}e^{-i\frac{\pi}{4}}\tilde{l}^{i\omega/\kappa_{+}}e^{-i\omega r_{+}}\Gamma\left(1-i\omega/\kappa_{+}\right)e^{-\frac{\pi\omega}{2\kappa_{+}}}
\]
which are the expressions given in Eqs. \eqref{ap}, \eqref{am},
\eqref{pIII1} and \eqref{pIII2}.

\end{document}